\documentclass[11pt]{article}
\usepackage{graphicx}
\usepackage{amsmath,amssymb,cite,times}
\usepackage{verbatim}

\DeclareMathOperator{\diag}{diag}

\newcommand{\bg}[1]{\boldsymbol{#1}} 
\newcommand{\bm}[1]{\mathbf{#1}} 

\newcommand\raiseT[2]{%
\setbox0\hbox{$#1{#2}$}\raise\dp0\box0}
\usepackage{algorithm,algorithmic}
\makeatletter
\newcommand{\ALOOP}[1]{\ALC@it\algorithmicloop\ #1%
  \begin{ALC@loop}}
\newcommand{\ENDALOOP}{\end{ALC@loop}\ALC@it\algorithmicendloop}

\makeatother

\title{\LARGE\textbf{Global spectral graph wavelet signature for surface analysis of carpal bones}}
\author{Majid Masoumi and A. Ben Hamza\\
Concordia Institute for Information Systems Engineering\\
Concordia University, Montreal, QC, Canada\\ [1ex]
Phone/fax: (514) 848-2424 \#5383 / (514) 848-3171 \\
E-mail: hamza@ciise.concordia.ca}
\date{}

\begin{document}
\maketitle

\begin{abstract}
In this paper, we present a spectral graph wavelet approach for shape analysis of carpal bones of human wrist. We apply a metric called global spectral graph wavelet signature for representation of cortical surface of the carpal bone based on eigensystem of Laplace-Beltrami operator. Furthermore, we propose a heuristic and efficient way of aggregating local descriptors of a carpal bone surface to global descriptor. The resultant global descriptor is not only isometric invariant, but also much more efficient and requires less memory storage. We perform experiments on shape of the carpal bones of ten women and ten men from a publicly-available database. Experimental results show the excellency of the proposed GSGW compared to recent proposed GPS embedding approach for comparing shapes of the carpal bones across populations.
\end{abstract}

\newpage
\section{Introduction}
Detecting unique phenotypes across populations can be achieved by quantitative analysis of bone shape, provided that the databases of normal and abnormal pathologies are available. The recent surge of interest in the spectral analysis of the Laplace-Beltrami operator (LBO) has resulted in a considerable number of spectral shape signatures that have been successfully applied to a broad range of areas, including manifold learning~\cite{Belkin:06}, object recognition and deformable shape analysis~\cite{Levy:06,Reuter:06,Rustamov:07,Bronstein:11,Masoumi:16,Rodola:SHREC17}, medical imaging~\cite{Chaudhari:14}, multimedia protection~\cite{Tarmissi:09}, and shape classification~\cite{Masoumi:17}. The diversified nature of these applications is a powerful testimony of the practical usage of spectral shapes signatures, which are usually defined as feature vectors representing local and/or global characteristics of a shape and may be broadly classified into two main categories: local and global descriptors. Local descriptors (also called point signatures) are defined on each point of the shape and often represent the local structure of the shape around that point, while global descriptors are usually defined on the entire shape.

Most point signatures may easily be aggregated to form global descriptors by integrating over the entire shape. Rustamov~\cite{Rustamov:07} proposed a local feature descriptor referred to as the global point signature (GPS), which is a vector whose components are scaled eigenfunctions of the LBO evaluated at each surface point. The GPS signature is invariant under isometric deformations of the shape, but it suffers from the problem of eigenfunctions' switching whenever the associated eigenvalues are close to each other. This problem was lately well handled by the heat kernel signature (HKS)~\cite{Sun:09}, which is a temporal descriptor defined as an exponentially-weighted combination of the LBO eigenfunctions. HKS is a local shape descriptor that has a number of desirable properties, including robustness to small perturbations of the shape, efficiency and invariance to isometric transformations. The idea of HKS was also independently proposed by G\c ebal~\textit{et al.}~\cite{Gebal:09} for 3D shape skeletonization and segmentation under the name of auto diffusion function. From the graph Fourier perspective, it can be seen that HKS is highly dominated by information from low frequencies, which correspond to macroscopic properties of a shape. To give rise to substantially more accurate matching than HKS, the wave kernel signature (WKS)~\cite{Aubry:11} was proposed as an alternative in an effort to allow access to high-frequency information. Using the Fourier transform's magnitude, Kokkinos~\textit{et al.}~\cite{Kokkinos:10} introduced the scale invariant heat kernel signature (SIHKS), which is constructed based on a logarithmically sampled scale-space.

One of the simplest spectral shape signatures is Shape-DNA~\cite{Reuter:06}, which is an isometry-invariant global descriptor defined as a truncated sequence of the LBO eigenvalues arranged in increasing order of magnitude. Gao~\textit{et al.}~\cite{Gao:14} developed a variant of Shape-DNA, referred to as compact Shape-DNA (cShape-DNA), which is an isometry-invariant signature resulting from applying the discrete Fourier transform to the area-normalized eigenvalues of the LBO. Chaudhari~\textit{et al.}~\cite{Chaudhari:14} presented a slightly modified version of the GPS signature by setting the LBO eigenfunctions to unity. This signature, called GPS embedding, is defined as a truncated sequence of inverse square roots of the area-normalized eigenvalues of the LBO. A comprehensive list of spectral descriptors can be found in~\cite{Lian:13}.

On the other hand, wavelet analysis has some major advantages over Fourier transform, which makes it an interesting alternative for many applications. In particular, unlike the Fourier transform, wavelet analysis is able to perform local analysis and also makes it possible to perform a multiresolution analysis. Classical wavelets are constructed by translating and scaling a mother wavelet, which is used to generate a set of functions through the scaling and translation operations. The wavelet transform coefficients are then obtained by taking the inner product of the input function with the translated and scaled waveforms. The application of wavelets to graphs (or triangle meshes in geometry processing) is, however, problematic and not straightforward due in part to the fact that it is unclear how to apply the scaling operation on a signal (or function) defined on the mesh vertices. To tackle this problem, Coifman~\textit{et al.}~\cite{Coifman:06} introduced the diffusion wavelets, which generalize the classical wavelets by allowing for multiscale analysis on graphs. The construction of diffusion wavelets interacts with the underlying graph through repeated applications of a diffusion operator, which induces a scaling process. Hammond~\emph{et al.}~\cite{Hammond:11} showed that the wavelet transform can be performed in the graph Fourier domain, and proposed a spectral graph wavelet transform that is defined in terms of the eigensystem of the graph Laplacian matrix. More recently, a spectral graph wavelet signature (SGWS) was introduced in~\cite{Chunyuan:13b}, and it has shown superior performance over HKS and WKS in 3D shape retrieval. SGWS is a multiresolution local descriptor that is not only isometric invariant, but also compact, easy to compute and combines the advantages of both band-pass and low-pass filters.

A popular approach for transforming local descriptors into global representations that can be used for surface analysis is the bag-of-features (BoF) paradigm~\cite{Bronstein:11}. The BoF model represents each object in the dataset as a collection of unordered feature descriptors extracted from local areas of the shape, just as words are local features of a document. A baseline BoF approach quantizes each local descriptor to its nearest cluster center using K-means clustering and then encodes each shape as a histogram over cluster centers by counting the number of assignments per cluster. These cluster centers form a visual vocabulary or codebook whose elements are often referred to as visual words or codewords. Although the BoF paradigm has been shown to provide significant levels of performance, it does not, however, take into consideration the spatial relations between features, which may have an adverse effect not only on its descriptive ability but also on its discriminative power. To account for the spatial relations between features, Bronstein~\textit{et al.} introduced a generalization of a bag of features, called spatially sensitive bags of features (SS-BoF)~\cite{Bronstein:11}. The SS-BoF is a global descriptor defined in terms of mid-level features and the heat kernel, and can be represented by a square matrix whose elements represent the frequency of appearance of nearby codewords in the vocabulary. In the same spirit, Bu~\textit{et al.}~\cite{Bu:14} recently proposed the geodesic-aware bags of features (GA-BoF) for 3D shape classification by replacing the heat kernel in SS-BoF with a geodesic exponential kernel.

In this paper, we introduce a global spectral graph wavelet framework that characterizes the shape of the cortical surface of a carpal bone in terms of eigensystem of Laplace-Beltrami operator. In a bid to aggregate the local descriptors and transform them into global representation we multiply the global spectral graph wavelet matrix of each shape by its area matrix. The resultant global descriptor is not only isometric invariant, but also much more efficient and requires less memory storage. We prove through experiment on publicly-available database \cite{Moore:07} that our proposed GSGW approach yields better performance in terms of MANOVA and permutation test compared to existing methods.

The rest of this paper is organized as follows. In Section \ref{background}, we briefly overview the Laplace-Beltrami operator. In Section \ref{method}, we introduce a global spectral graph wavelet framework, and we discuss in detail its main algorithmic steps. Experimental results are presented in Section \ref{experiment}. Finally, we conclude in Section \ref{conclusion} and point out some future work directions.
\section{Background} \label{background}
A 3D shape is usually modeled as a triangle mesh $\mathbb{M}$ whose vertices are sampled from a Riemannian manifold. A triangle mesh $\mathbb{M}$ may be defined as a graph $\mathbb{G}=(\mathcal{V},\mathcal{E})$ or $\mathbb{G}=(\mathcal{V},\mathcal{T})$, where $\mathcal{V}=\{\bm{v}_{1},\ldots,\bm{v}_{m}\}$ is the set of vertices, $\mathcal{E}=\{e_{ij}\}$ is the set of edges, and $\mathcal{T}=\{\bm{t}_{1},\ldots,\bm{t}_{g}\}$ is the set of triangles. Each edge $e_{ij}=[\bm{v}_{i},\bm{v}_{j}]$ connects a pair of vertices $\{\bm{v}_{i},\bm{v}_{j}\}$. Two distinct vertices $\bm{v}_{i},\bm{v}_{j}\in {\cal V}$ are adjacent (denoted by $\bm{v}_{i}\sim\bm{v}_{j}$ or simply $i\sim j$) if they are connected by an edge, i.e. $e_{ij}\in {\cal E}$.
\subsection{Laplace-Beltrami Operator}
Given a compact Riemannian manifold $\mathbb{M}$, the space $L^{2}(\mathbb{M})$ of all smooth, square-integrable functions on $\mathbb{M}$ is a Hilbert space endowed with inner product
$\langle f_{1},f_{2} \rangle=\int_{\mathbb{M}} f_{1}(\bm{x}) f_{2}(\bm{x})\,da(\bm{x})$, for all $f_{1}, f_{2}\in L^{2}(\mathbb{M})$, where $da(x)$ (or simply $dx$) denotes the measure from the area element of a Riemannian metric on $\mathbb{M}$. Given a twice-differentiable, real-valued function $f:\,\mathbb{M}\to\mathbb{R}$, the Laplace-Beltrami operator (LBO) is defined as $\Delta_{\mathbb{M}} f=-\mathrm{div}(\nabla_{\mathbb{M}} f)$, where $\nabla_{\mathbb{M}} f$ is the intrinsic gradient vector field and $\mathrm{div}$ is the divergence operator~\cite{Krim:15,Rosenberg:97}. The LBO is a linear, positive semi-definite operator acting on the space of real-valued functions defined on $\mathbb{M}$, and it is a generalization of the Laplace operator to non-Euclidean spaces.

\medskip
\noindent{\textbf{Discretization}}\quad A real-valued function $f:\mathcal{V}\to\mathbb{R}$ defined on the mesh vertex set may be represented as an $m$-dimensional vector $\bm{f}=(f(i))\in\mathbb{R}^{m}$, where the $i$th component $f(i)$ denotes the function value at the $i$th vertex in $\mathcal{V}$. Using a mixed finite element/finite volume method on triangle meshes~\cite{Meyer:03}, the value of $\Delta_{\mathbb{M}}f$ at a vertex $\bm{v}_{i}$ (or simply $i$) can be approximated using the \textrm{cotangent weight} scheme as follows:
\begin{equation}
\Delta_{\mathbb{M}}f(i)\approx\frac{1}{a_{i}}\sum_{j\sim i}
\frac{\cot\alpha_{ij} + \cot\beta_{ij}}{2}\bigl(f(i)-f(j)\bigr),
\label{Eq:LapBeltrOperator}
\end{equation}
where $\alpha_{ij}$ and $\beta_{ij}$ are the angles $\angle(\bm{v}_{i}\bm{v}_{k_1}\bm{v}_{j})$ and $\angle(\bm{v}_{i}\bm{v}_{k_2}\bm{v}_{j})$ of two faces $\bm{t}^{\alpha}=\{\bm{v}_{i},\bm{v}_{j},\bm{v}_{k_1}\}$ and $\bm{t}^{\beta}=\{\bm{v}_{i},\bm{v}_{j},\bm{v}_{k_2}\}$ that are adjacent to the edge $[i,j]$, and $a_i$ is the area of the Voronoi cell at vertex $i$. It should be noted that the cotangent weight scheme is numerically consistent and preserves several important properties of the continuous LBO, including symmetry and positive semi-definiteness~\cite{Wardetzky:07}.

\begin{figure}[!htb]
\setlength{\tabcolsep}{.5em}
\centering
\begin{tabular}{cccc}
\includegraphics[scale=.17]{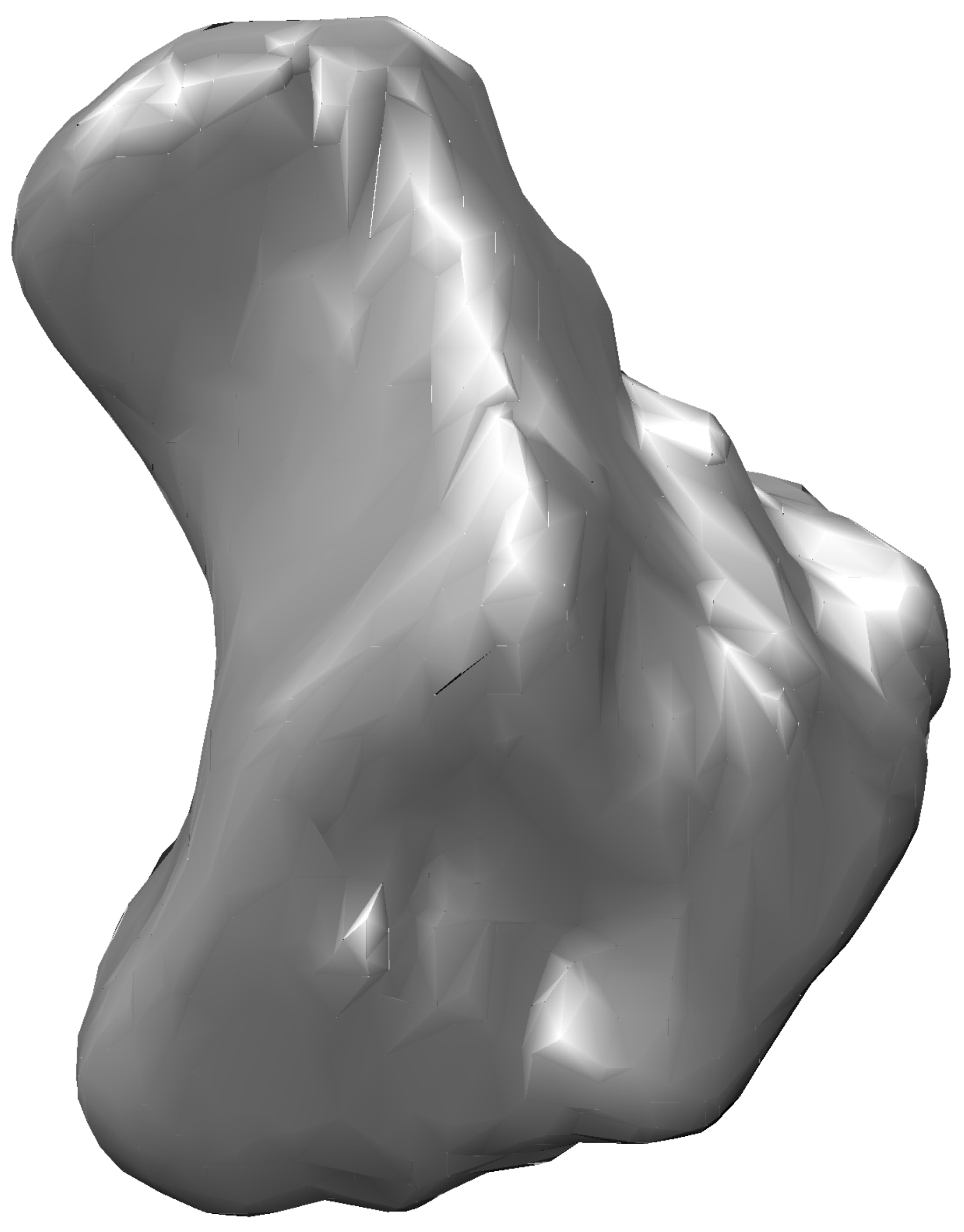}&
\includegraphics[scale=.17]{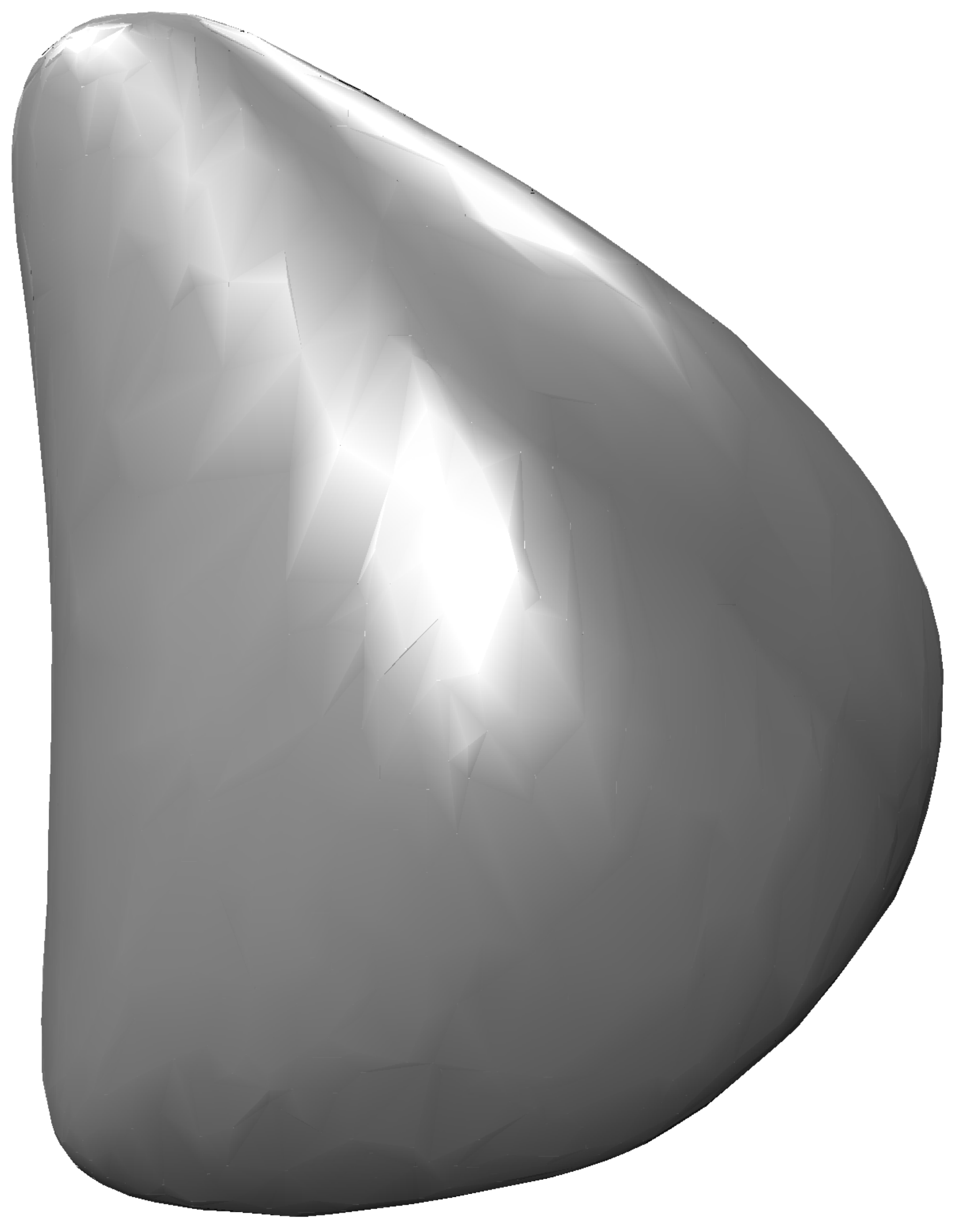}&
\includegraphics[scale=.17]{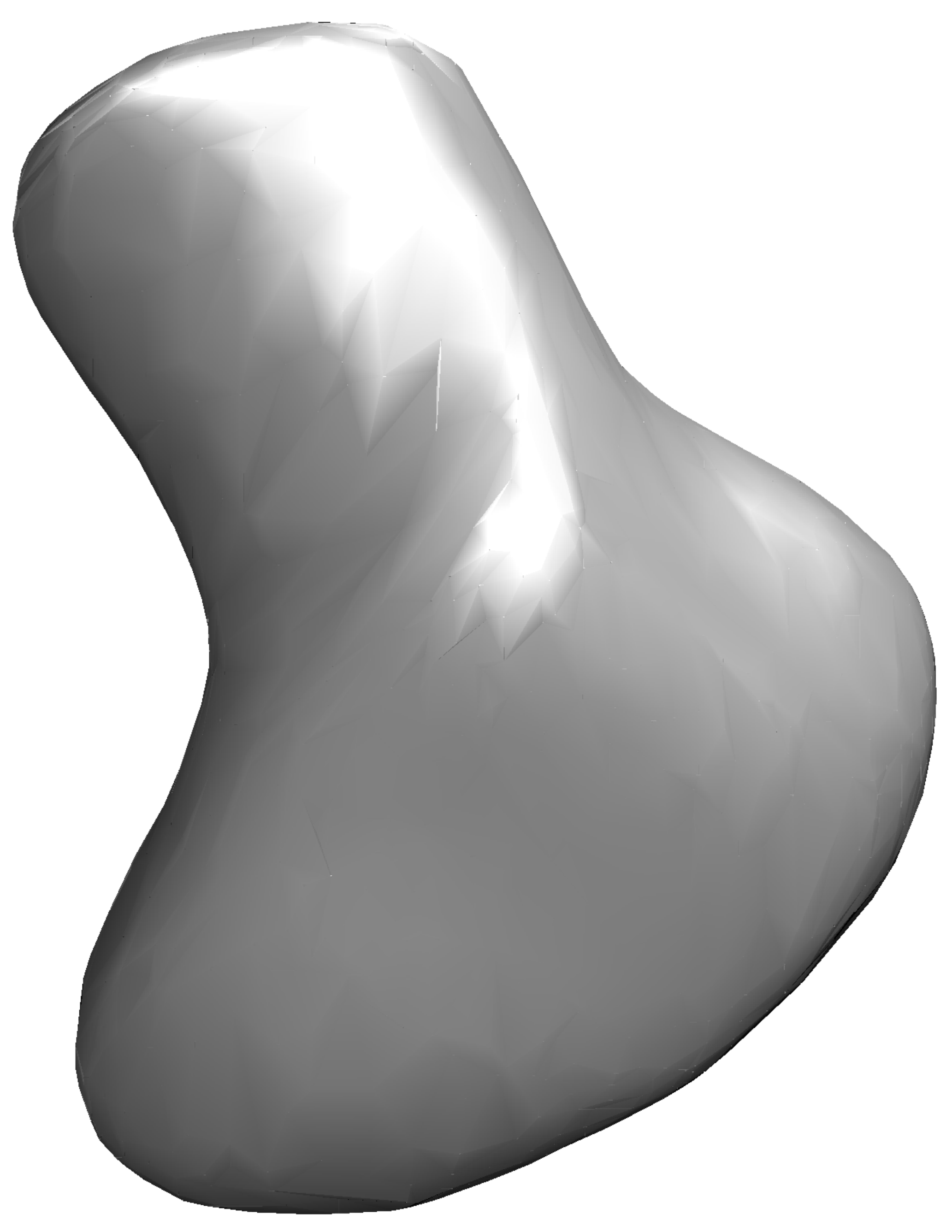}&
\includegraphics[scale=.17]{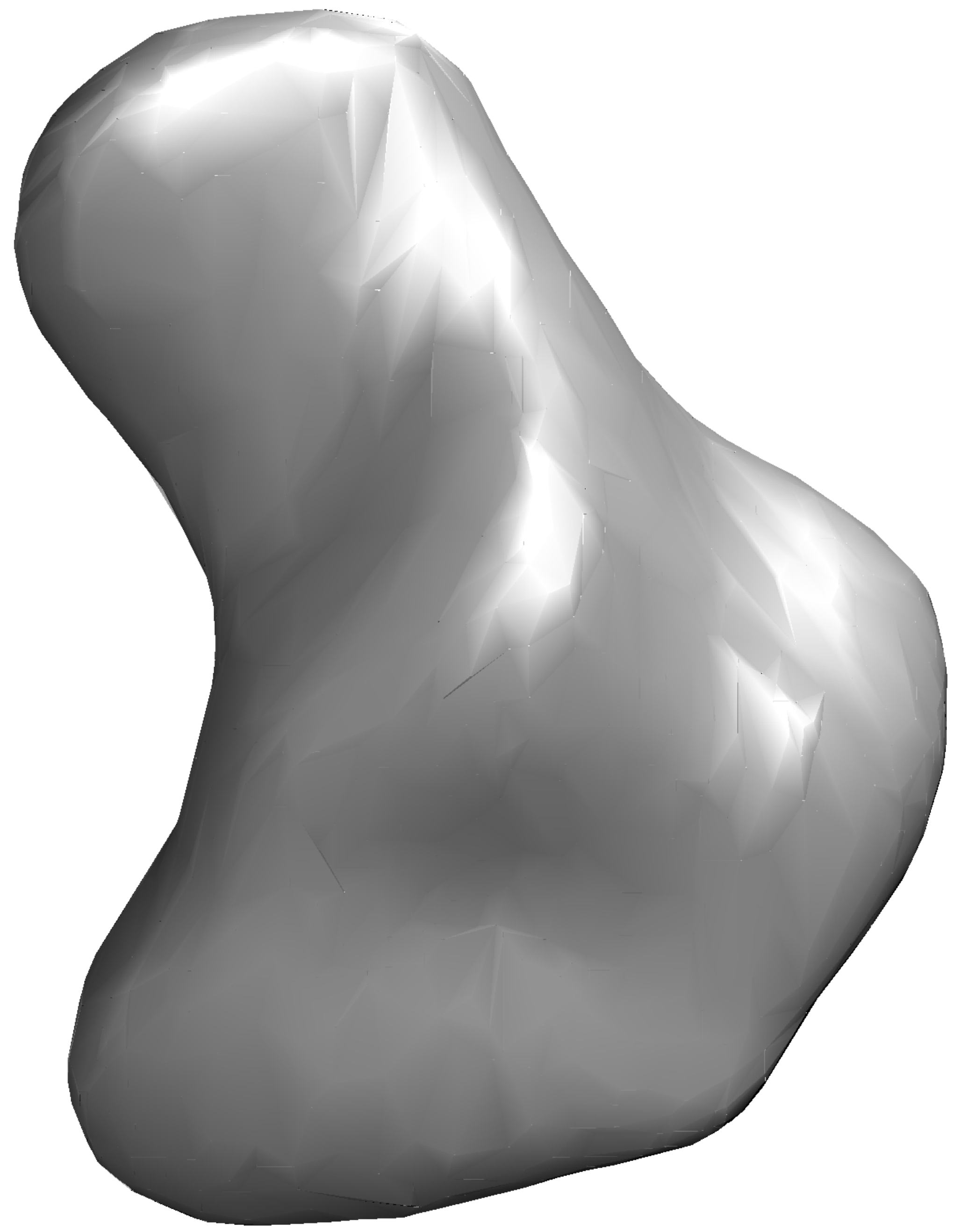}\\
(a) & (b) & (c) & (d)
\end{tabular}
\caption{(a) 3D representation of hamate bone in a healthy man and visualization of the reconstructed hamate bone using the (b) fifth, (c) twentieth and (d) fiftieth eigenfunctions of hamate bone.}
\label{Fig:hamate}
\end{figure}

\medskip
\noindent{\textbf{Spectral Analysis}}\quad The $m\times m$ matrix associated to the discrete approximation of the LBO is given by $\bm{L}=\bm{A}^{-1}\bm{W}$, where $\bm{A}=\mathrm{diag}(a_{i})$ is a positive definite diagonal matrix (mass matrix), and $\bm{W}=\diag(\sum_{k\neq i} c_{ik})-(c_{ij})$ is a sparse symmetric matrix (stiffness matrix). Each diagonal element $a_i$ is the area of the Voronoi cell at vertex $i$, and the weights $c_{ij}$ are given by
\begin{equation}
c_{ij}=
\begin{cases}
\dfrac{\cot\alpha_{ij} + \cot\beta_{ij}}{2} &\mbox{if } i\sim j \\
0 & \mbox{o.w.}
\end{cases}
\end{equation}
where $\alpha_{ij}$ and $\beta_{ij}$ are the opposite angles of two triangles that are adjacent to the edge $[i,j]$.

The eigenvalues and eigenvectors of $\bm{L}$ can be found by solving the generalized eigenvalue problem $\bm{W}\bg{\varphi}_{\ell}=\lambda_{\ell}\bm{A}\bg{\varphi}_{\ell}$ using for instance the Arnoldi method of ARPACK, where $\lambda_{\ell}$ are the eigenvalues and $\bg{\varphi}_{\ell}$ are the unknown associated eigenfunctions (i.e. eigenvectors which can be thought of as functions on the mesh vertices). We may sort the eigenvalues in ascending order as $0=\lambda_1<\lambda_{2}\le\dots\le\lambda_{m}$ with associated orthonormal eigenfunctions $\bg{\varphi}_{1}, \bg{\varphi}_{2},\dots,\bg{\varphi}_{m}$, where the orthogonality of the eigenfunctions is defined in terms of the $\bm{A}$-inner product, i.e.
\begin{equation}
\langle\bg{\varphi}_{k},\bg{\varphi}_{\ell}\rangle_{\bm{A}}=
\sum_{i=1}^{m}a_{i}\varphi_{k}(i)\varphi_{\ell}(i)=\delta_{k\ell},
\quad\text{for all } k, \ell=1,\dots,m.
\label{eq:orthoeig}
\end{equation}
We may rewrite the generalized eigenvalue problem in matrix form as $\bm{W}\bm{\Phi}=\bm{A}\bm{\Phi}\bm{\Lambda}$, where $\bm{\Lambda}=\diag(\lambda_{1},\dots,\lambda_{m})$ is an $m\times m$ diagonal matrix with the $\lambda_{\ell}$ on the diagonal, and $\bm{\Phi}$ is an $m\times m$ orthogonal matrix whose $\ell$th column is the unit-norm eigenvector $\bg{\varphi}_{\ell}$.

The ability of eigenvectors of LBO for rendering the shape-based features is depicted in Figure~\ref{Fig:hamate}. As shown, the lower-order eigenvectors capture the global structure of shape, while by increasing the number of eigenvectors more details of the curvature of the bone are captured.

\begin{figure}[!htb]
\setlength{\tabcolsep}{.5em}
\centering
\begin{tabular}{cccc}
\includegraphics[scale=.16]{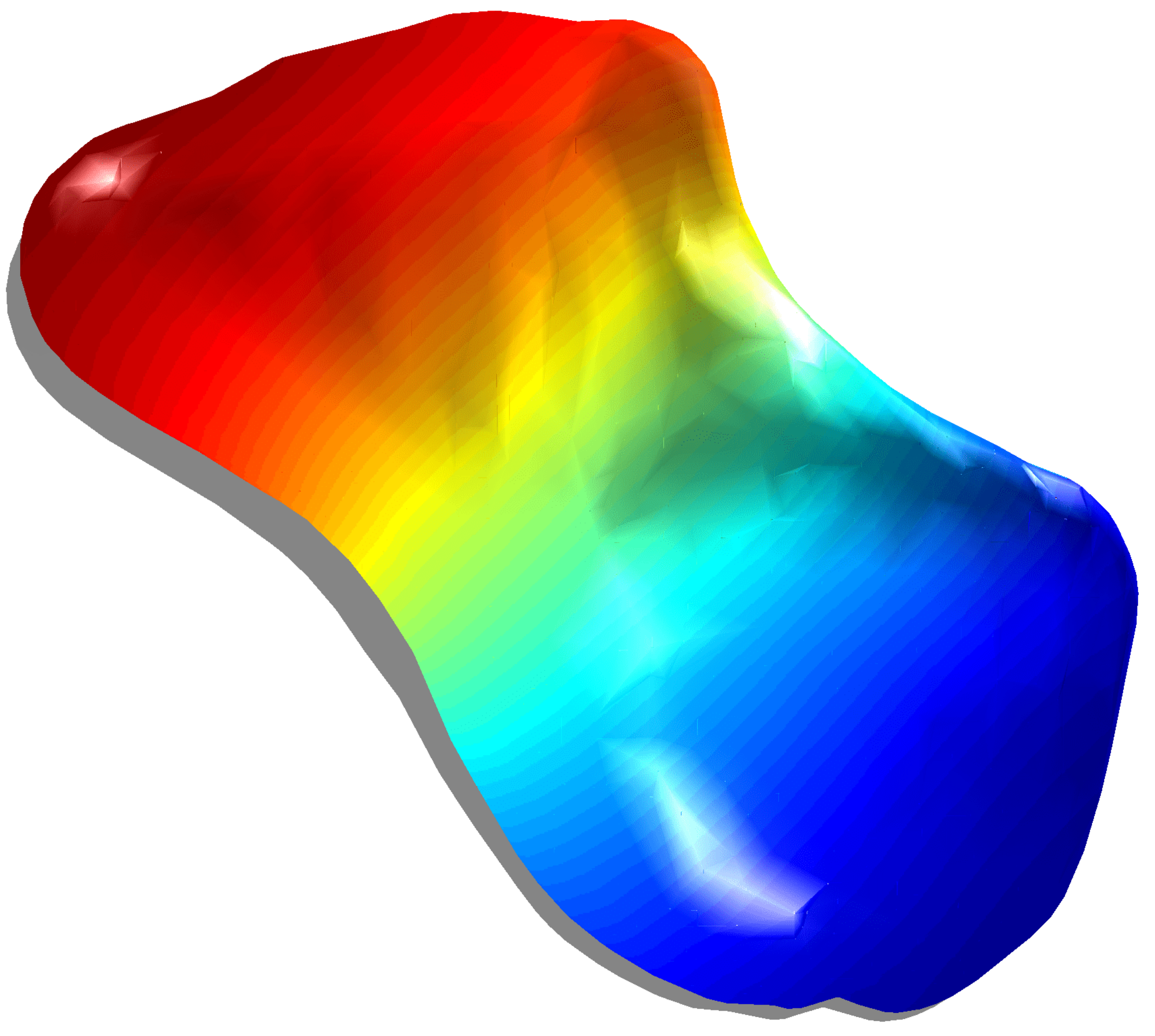}&
\includegraphics[scale=.16]{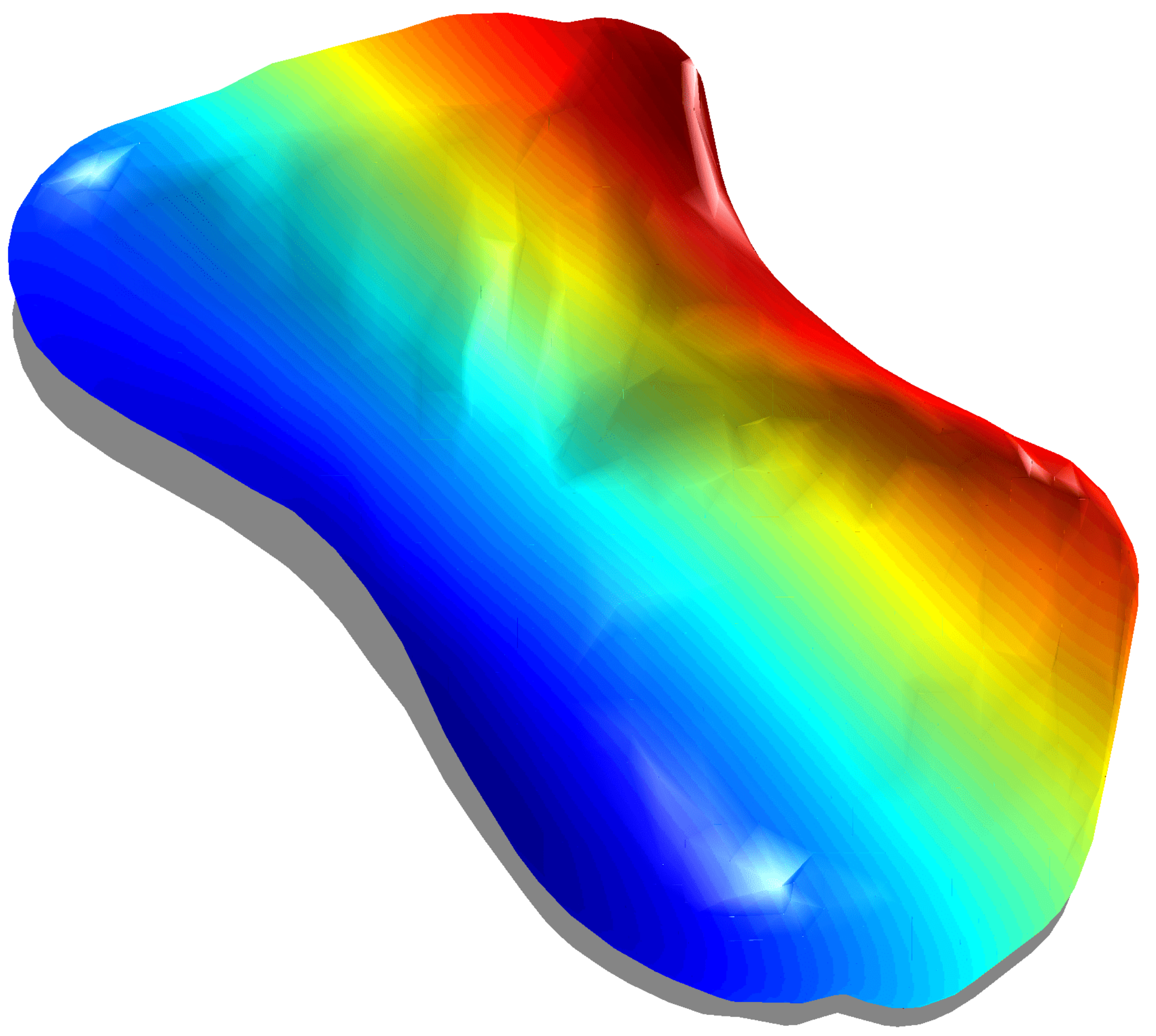}&
\includegraphics[scale=.16]{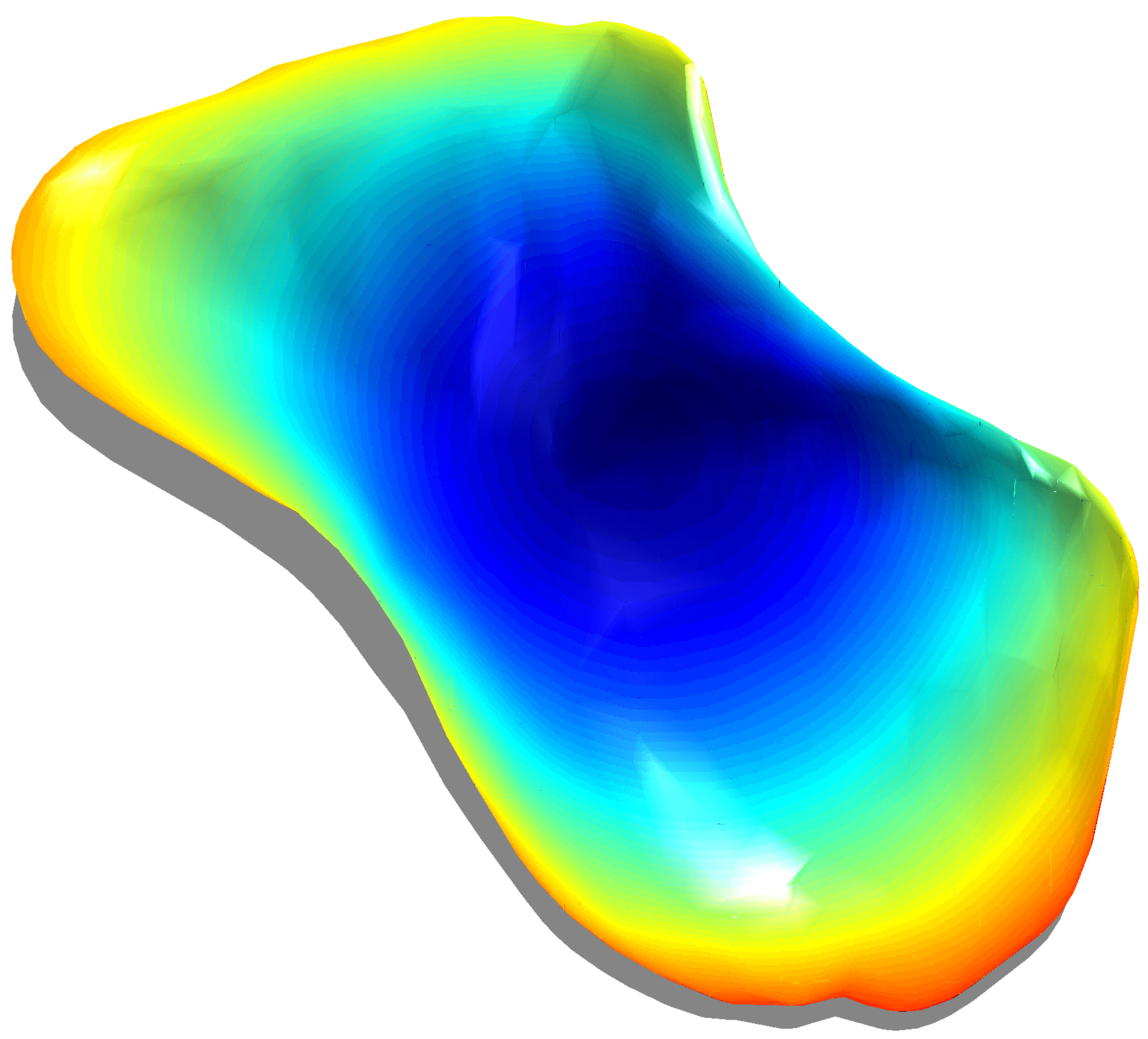}&
\includegraphics[scale=.16]{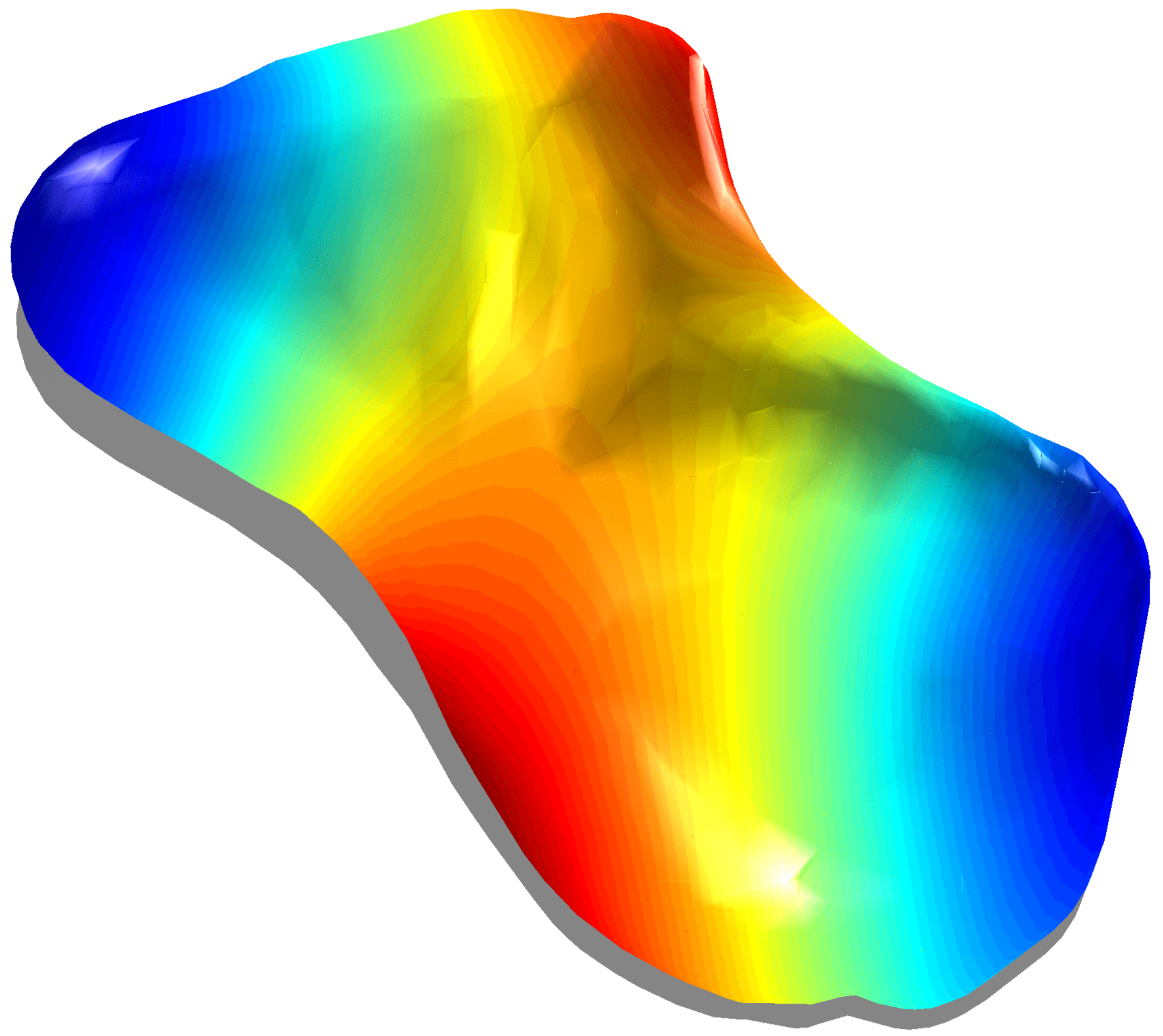}\\
(a) & (b) & (c) & (d)
\end{tabular}
\caption{Surface representation of the scaphoid bone showing the distribution of the (a) second, (b) third, (c) fourth and (d) fifth eigenfunction of the LBO mapped to the bone surface.}
\label{Fig:hamateEigen}
\end{figure}

The successful use of the LBO eigenvalues and eigenfunctions in shape analysis is largely attributed due to their isometry invariance and robustness to noise. Moreover, the eigenfunctions associated to the smallest eigenvalues capture well the large-scale properties of a shape. As shown in Figure~\ref{Fig:hamateEigen}, the (non-trivial) eigenfunctions of the LBO encode important information about the intrinsic global geometry of a shape. Notice that the eigenfunctions associated with larger eigenvalues oscillate more rapidly. Blue regions indicate negative values of the eigenfunctions and red colors regions indicate positive values, while green and yellow regions in between.

\begin{figure}[!htb]
\setlength{\tabcolsep}{.5em}
\centering
\begin{tabular}{c}
\includegraphics[scale=.6]{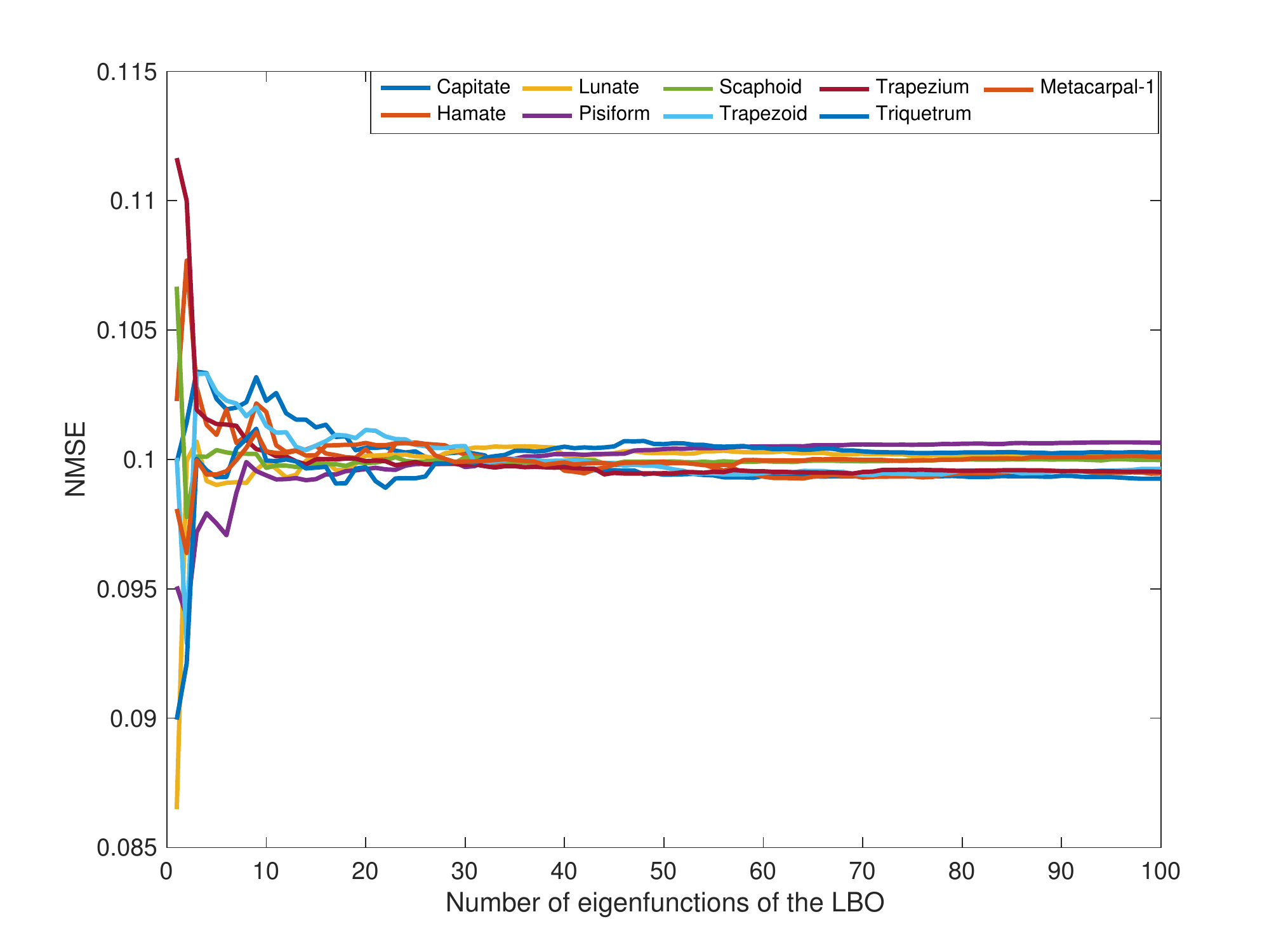}
\end{tabular}
\caption{The normalized mean square error between original carpal bone and the reconstructed carpal bone achieved by the first 100 eigenfunctions of LBO for a female.}
\label{Fig:NMSE}
\end{figure}

Figure \ref{Fig:NMSE} shows the normalized mean-squared error between the original bone surface and that reconstructed from an increasing number of eigenvectors of the LBO. As can be seen, just a small number of eigenvectors, i.e. $\ell\in (20,30)$, is sufficient to capture well features on the carpal bone surface to analyze shape differences in population study.
Rendering carpal bone surface in lower-dimension enjoys some advantages like not being vulnerable to tessellation noise or image segmentation.

\section{Method}\label{method}
In this section, we provide a detailed description of our GSGW  framework for analysis of cortical surface of a carpal bone using spectral graph wavelets. We start by defining the spectral graph wavelet transforms on a Riemannian manifold. We show how to build local descriptors from spectral graph wavelets and its subcomponent functions. Then, we propose a novel method to aggregate local descriptors to make global ones. Finally, we provide the main algorithmic steps of our carpal bone analysis framework.

\medskip

\subsection{Local Descriptors}
Wavelets are useful in describing functions at different levels of resolution. To characterize the localized context around a mesh vertex $j\in\mathcal{V}$, we assume that the signal on the mesh is a unit impulse function, that is $f(i)=\delta_j(i)$ at each mesh vertex $i\in\mathcal{V}$. The spectral graph wavelet coefficients are expressed as
\begin{equation}
W_{\delta_j}(t,j)=\langle \bg{\delta}_{j},\bg{\psi}_{t,j} \rangle=\sum_{\ell=1}^{m}a_{j}^{2}g(t\lambda_\ell)\varphi_{\ell}^{2}(j),
\label{DeltaW_coefficients}
\end{equation}
and that the coefficients of the scaling function are
\begin{equation}
S_{\delta_j}(j)=\sum_{\ell=1}^{m}a_{j}^{2} h(\lambda_\ell)\varphi_{\ell}^{2}(j).
 \label{DeltaS_coefficients}
\end{equation}
Following the multiresolution analysis, the spectral graph wavelet and scaling function coefficients are collected to form the the spectral graph wavelet signature at vertex $j$ as follows:
\begin{equation}
\bm{s}_j=\{\bm{s}_{L}(j)\mid L=1,\dots,R\},
 \label{Eq:SGWSignature}
\end{equation}
where $R$ is the resolution parameter, and $\bm{s}_{L}(j)$ is the shape signature at resolution level $L$ given by
\begin{equation}
\bm{s}_{L}(j)=\{W_{\delta_j}(t_k,j)\mid k=1,\dots,L\}\cup\{S_{\delta_j}(j)\}.
 \label{Eq:SGWSignatureLevel}
\end{equation}
The wavelet scales $t_k$ ($t_k > t_{k+1}$) are selected to be logarithmically equispaced between maximum and minimum scales $t_1$ and $t_L$, respectively. Thus, the resolution level $L$ determines the resolution of scales to modulate the spectrum. At resolution $R=1$, the spectral graph wavelet signature $\bm{s}_j$ is a 2-dimensional vector consisting of two elements: one element, $W_{\delta_j}(t_1,j)$, of spectral graph wavelet function coefficients and another element, $S_{\delta_j}(j)$, of scaling function coefficients. And at resolution $R=2$, the spectral graph wavelet signature $\bm{s}_j$ is a 5-dimensional vector consisting of five elements (four elements of spectral graph wavelet function coefficients and one element of scaling function coefficients). In general, the dimension of a spectral graph wavelet signature $\bm{s}_j$ at vertex $j$ can be expressed in terms of the resolution $R$ as follows:
\begin{equation}
p = \frac{(R+1)(R+2)}{2}-1.
\end{equation}
Hence, for a $p$-dimensional signature $\bm{s}_j$, we define a $p\times m$ spectral graph wavelet signature matrix as $\bm{S}=(\bm{s}_1,\dots,\bm{s}_m)$, where $\bm{s}_j$ is the signature at vertex $j$ and $m$ is the number of mesh vertices. In our implementation, we used the Mexican hat wavelet as a kernel generating function $g$. In addition, we used the scaling function $h$ given by
\begin{equation}
h(x)=\gamma \exp\left(-\left (\frac{x}{ 0.6\lambda_{\min}} \right)^4\right),
\end{equation}
where $\lambda_{\min}=\lambda_{\max}/20$ and $\gamma$ is set such that $h(0)$ has the same value as the maximum value of $g$. The maximum and minimum scales are set to $t_{1}=2/\lambda_{\min}$ and $t_{L}=2/\lambda_{\max}$.

\begin{figure}[htb]
\centering
\begin{tabular}{cc}
\includegraphics[scale=.5]{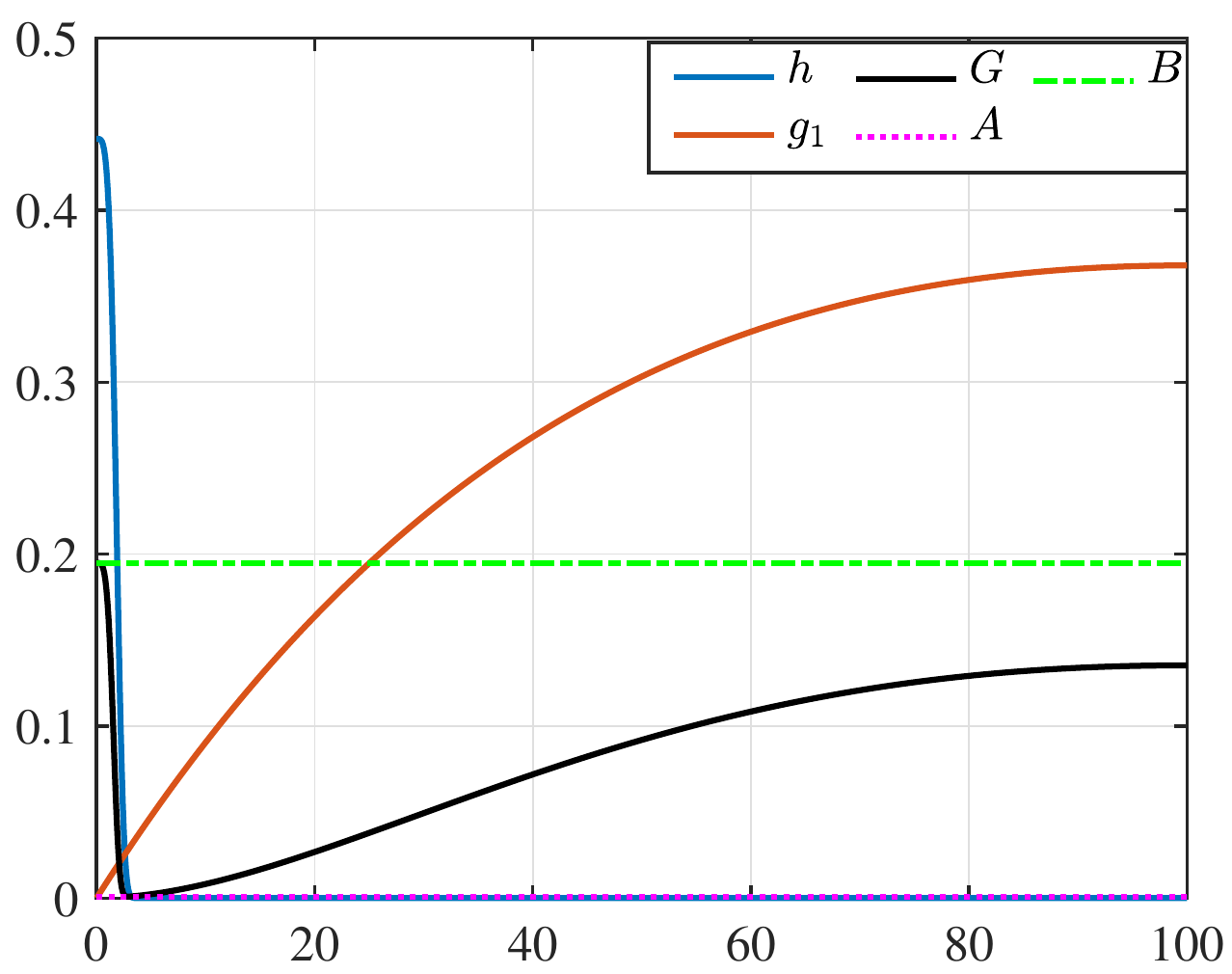}&
\includegraphics[scale=.5]{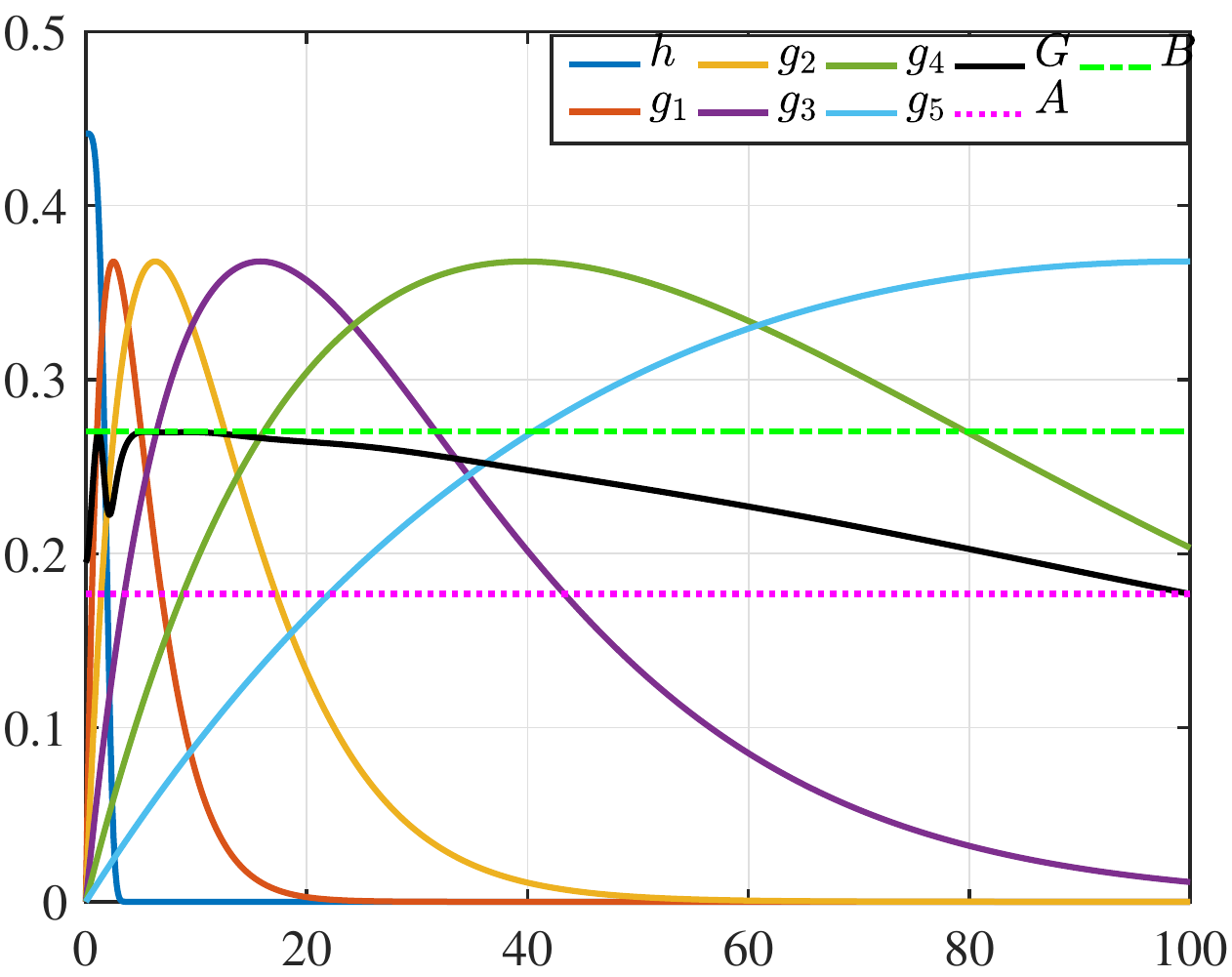}\\
(a) Mexican hat kernel for \emph{R}=1 & (b) Mexican hat kernel for \emph{R}=5\\
\includegraphics[scale=.5]{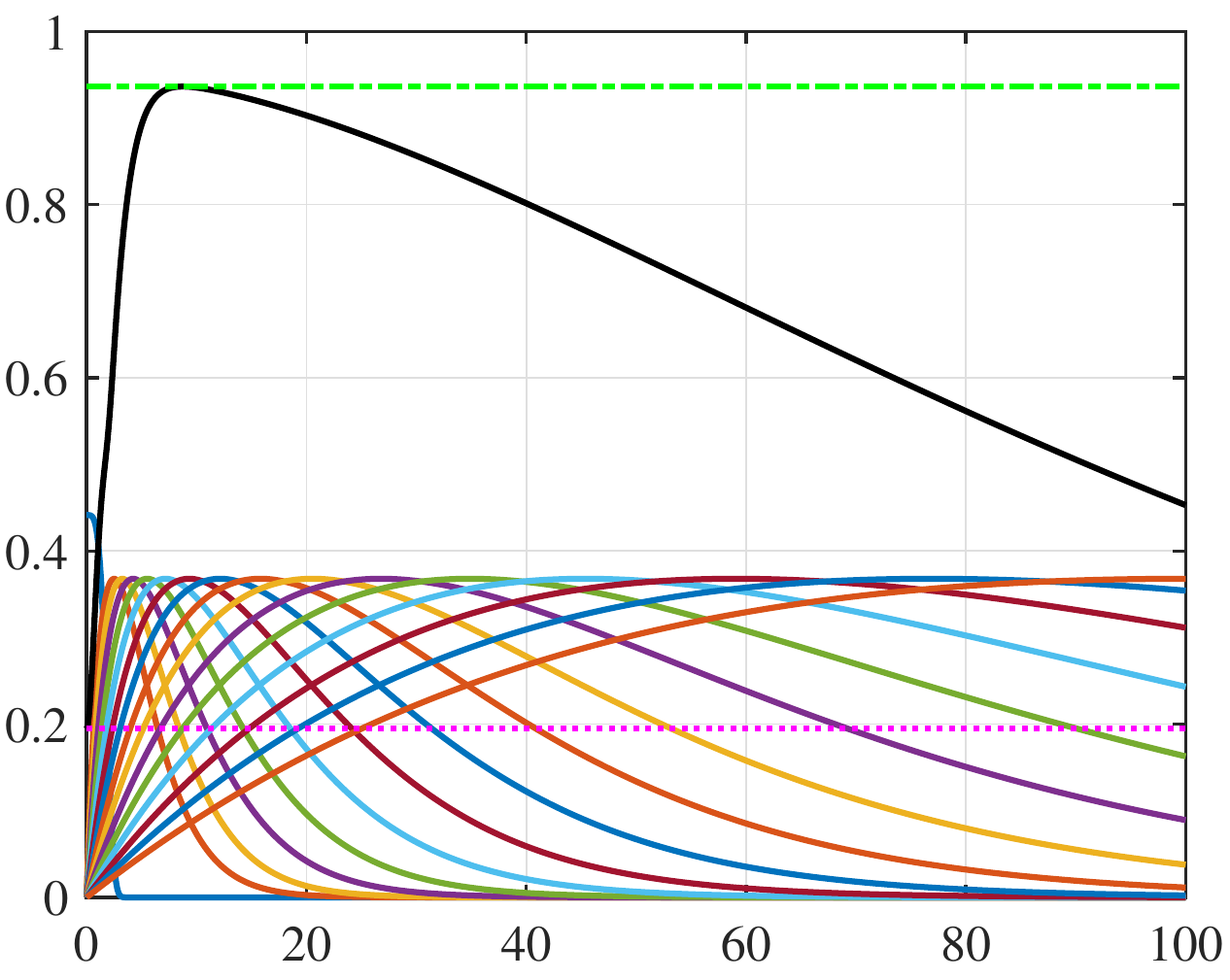}&
\includegraphics[scale=.5]{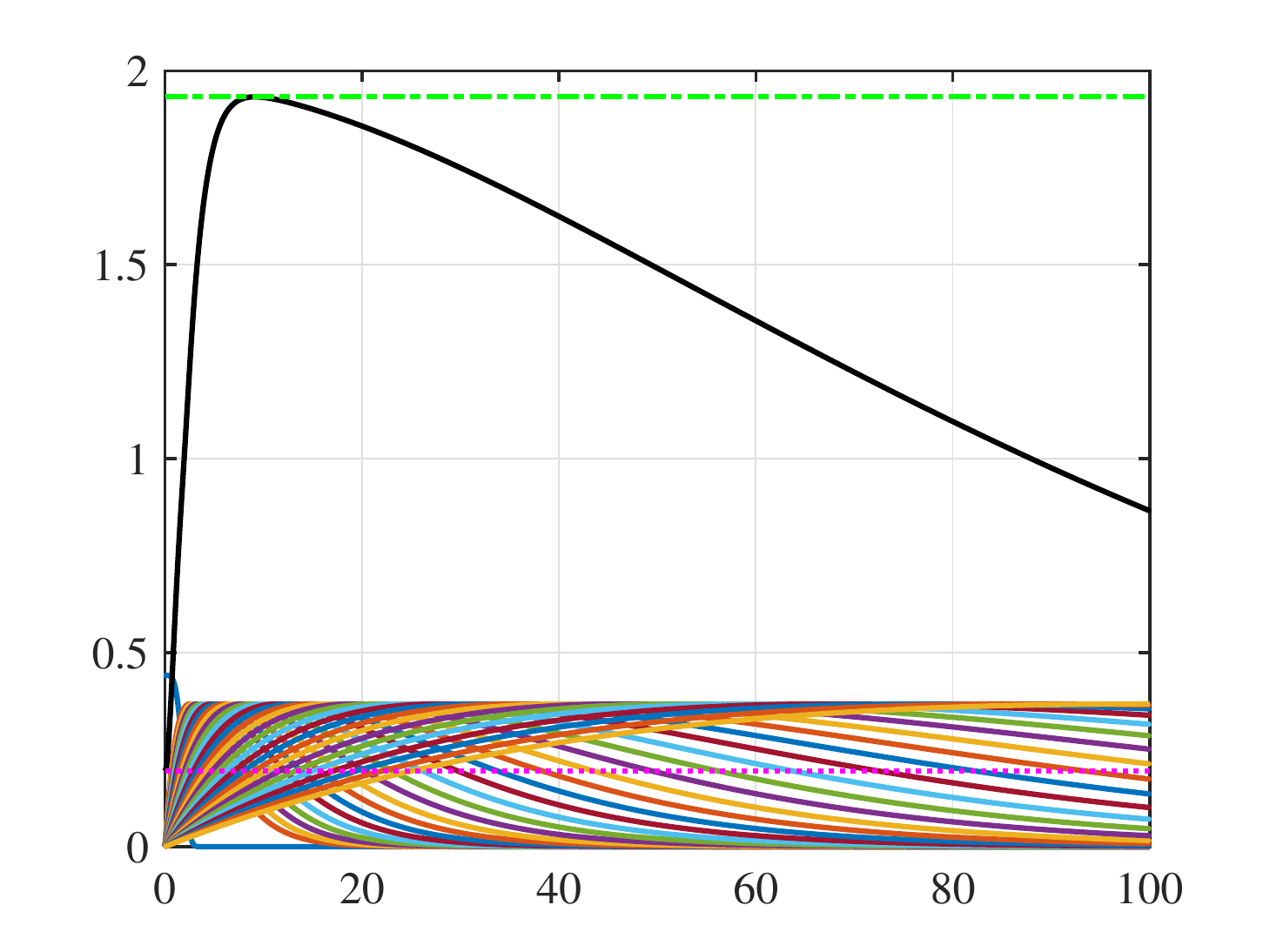}\\
(c) Mexican hat kernel for \emph{R}=15 & (d) Mexican hat kernel for \emph{R}=30\\
\end{tabular}
\caption{Spectrum modulation using different kernel functions at various resolutions. The dark line is the squared sum function $G$, while the dash-dotted and the dotted lines are upper and lower bounds ($B$ and $A$) of $G$, respectively.}
\label{Fig:SGWT}
\end{figure}

The geometry captured at each resolution $R$ of the spectral graph wavelet signature can be viewed as the area under the curve $G$ shown in Figure~\ref{Fig:SGWT}. For a given resolution $R$, we can understand the information from a specific range of the spectrum as its associated areas under $G$. As the resolution $R$ increases, the partition of spectrum becomes tighter, and thus a larger portion of the spectrum is highly weighted.

\subsection{Global Descriptors}
We refer to the $p$-dimensional vector $\bm{s}_j = ({s}_1(j),\dots,{s}_p(j))$ consisting of the first $p$ wavelet and scaling functions as the \emph{spectral graph wavelet point signature} at vertex $j$ . Hence, we may represent the shape $\mathbb{M}$ by an $m\times p$ spectral graph wavelet signature matrix as $\bm{S}=(\bm{s}_1,\dots,\bm{s}_m)^T$ of $m$ signatures, each of which is of length $p$ . In other words, the rows of $\bm{S}$ are data points and the columns are features.

In a bid to aggregate the local descriptors and build global descriptor to represent a carpal bone surface we have to exploit bag-of-features paradigm. A major drawback of the BoF model is that it only considers the distribution of the codewords and disregards all information about the spatial relations between features, and hence the descriptive ability and discriminative power of the BoF paradigm may be negatively impacted. In addition, the BoF process is time-consuming since it requires different steps such as constructing dictionary, feature coding and feature pooling. To circumvent this limitation, we represent the shape $\mathbb{M}$ by the $p\times 1$ vector $\bm{g}= \bm{S}\bm{a}$, where $\bm{a}=(a_{1},\dots,a_{m})^T$ is a $m\times 1$ area vector. We refer to $\bm{g}$ as the global spectral graph wavelet (GSGW) descriptor of the carpal bone surface. In addition to circumvent the BoF process, the GSGW descriptor enjoys a number of desirable properties including simplicity, compactness, invariance to isometric deformations, and computational feasibility. Moreover, our spectral graph wavelet signature combines the advantages of both band-pass and low-pass filters.

\subsection{Proposed Algorithm}
 Our proposed carpal bone analysis algorithm consists of two main steps. The first step is to represent each bone in the dataset by a spectral graph wavelet signature matrix, which is a feature matrix consisting of local descriptors. More specifically, let $\mathcal{D}$ be a dataset of $n$ carpal bones modeled by triangle meshes $\mathbb{M}_1,\dots,\mathbb{M}_n$. We represent each surface in the dataset $\mathcal{D}$ by a $p\times m$ spectral graph wavelet signature matrix $\bm{S}=(\bm{s}_{1},\dots,\bm{s}_{m})\in\mathbb{R}^{p\times m}$, where $\bm{s}_i$ is the $p$-dimensional local descriptor at vertex $i$ and $m$ is the number of mesh vertices.

\begin{figure}[!htb]
\centering
\begin{tabular}{cc}
\includegraphics[scale=.6]{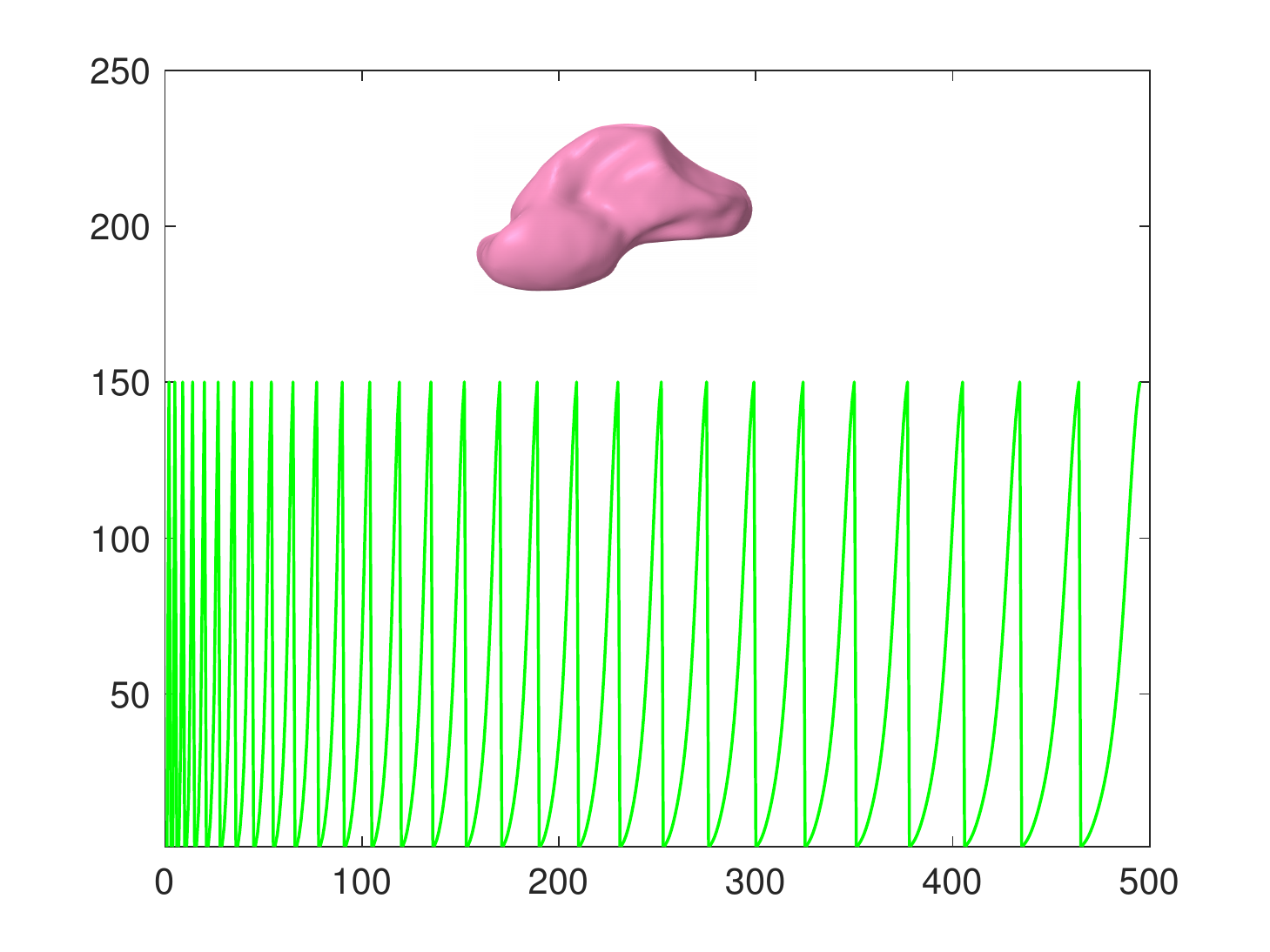} &
\includegraphics[scale=.6]{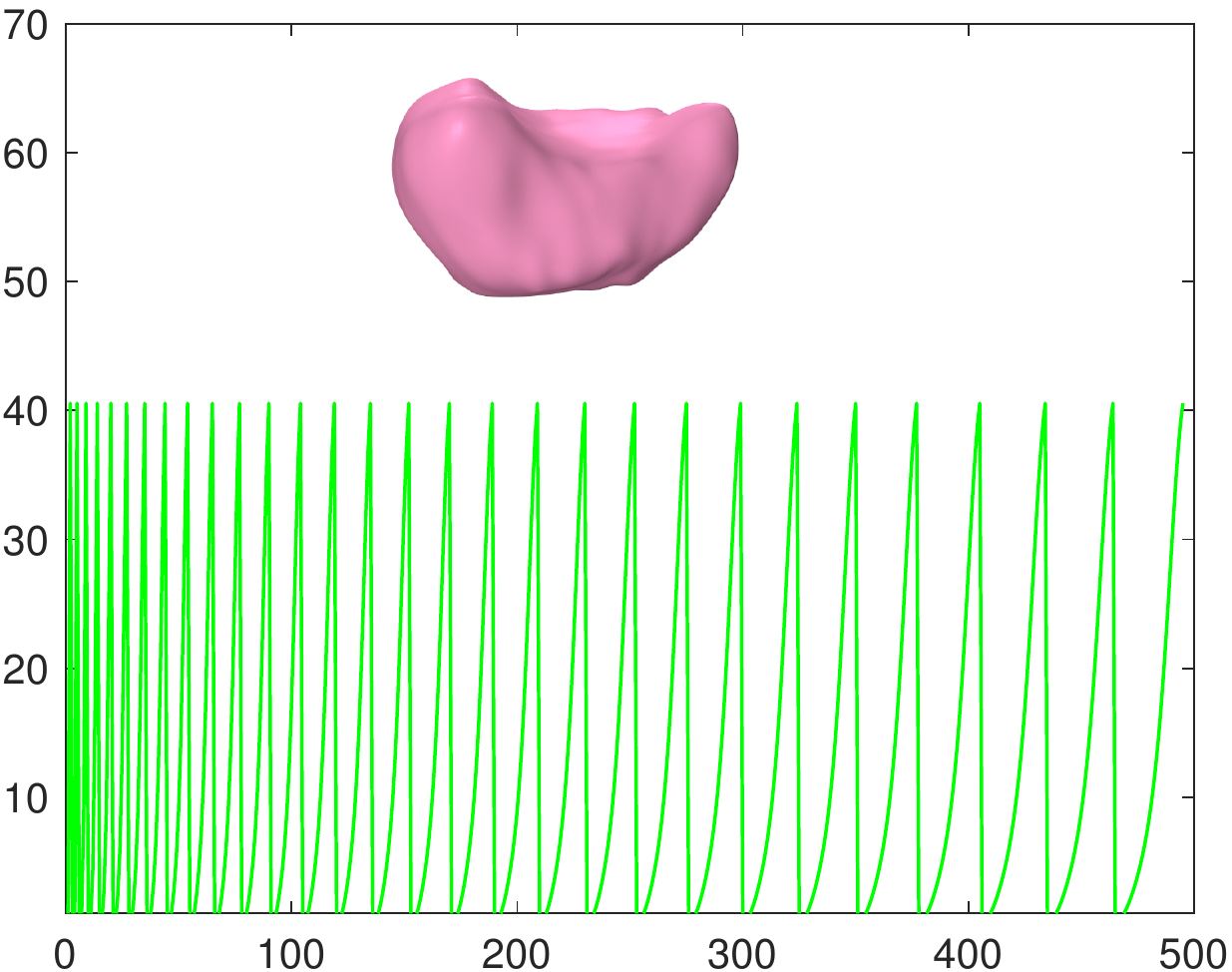}
\end{tabular}
\caption{Global spectral graph wavelet codes of two carpal bones i.e. capitate (left) and lunate (right).}
\label{Fig:GSGW1}
\end{figure}

In the second step, the spectral graph wavelet signatures $\bm{s}_{i}$ are aggregated to feature vector $\bm{g}$ using computing $\bm{g}_i= \bm{S}_i\bm{a}_i$ for each carpal bone $\mathbb{M}$. Subsequently, all feature vectors $\bm{x}_{i}$ of all $n$ shapes in the dataset are arranged into a $n\times p$ data matrix $\bm{X}=(\bm{x}_{1},\dots,\bm{x}_{n})^T$. Figure \ref{Fig:GSGW1} displays the GSGW codes of two carpal bones (capitate and lunate) from two different classes of wrist dataset \cite{Moore:07} .

To assess the performance of the proposed GSGW framework, we employed two commonly-used evaluation criteria, the MANOVA and permutation test, which will be discussed in more detail in the next section.

\bigskip

\section{Experimental Results}\label{experiment}
In this section, we evaluate the performance of our proposed GSGW approach for analysis of carpal bone surfaces through extensive experiments. The effectiveness of our method is validated by performing a comprehensive comparison with recent GPS approach \cite{Chaudhari:14}.

\begin{figure}[htb]
\centering
\includegraphics[scale=.3]{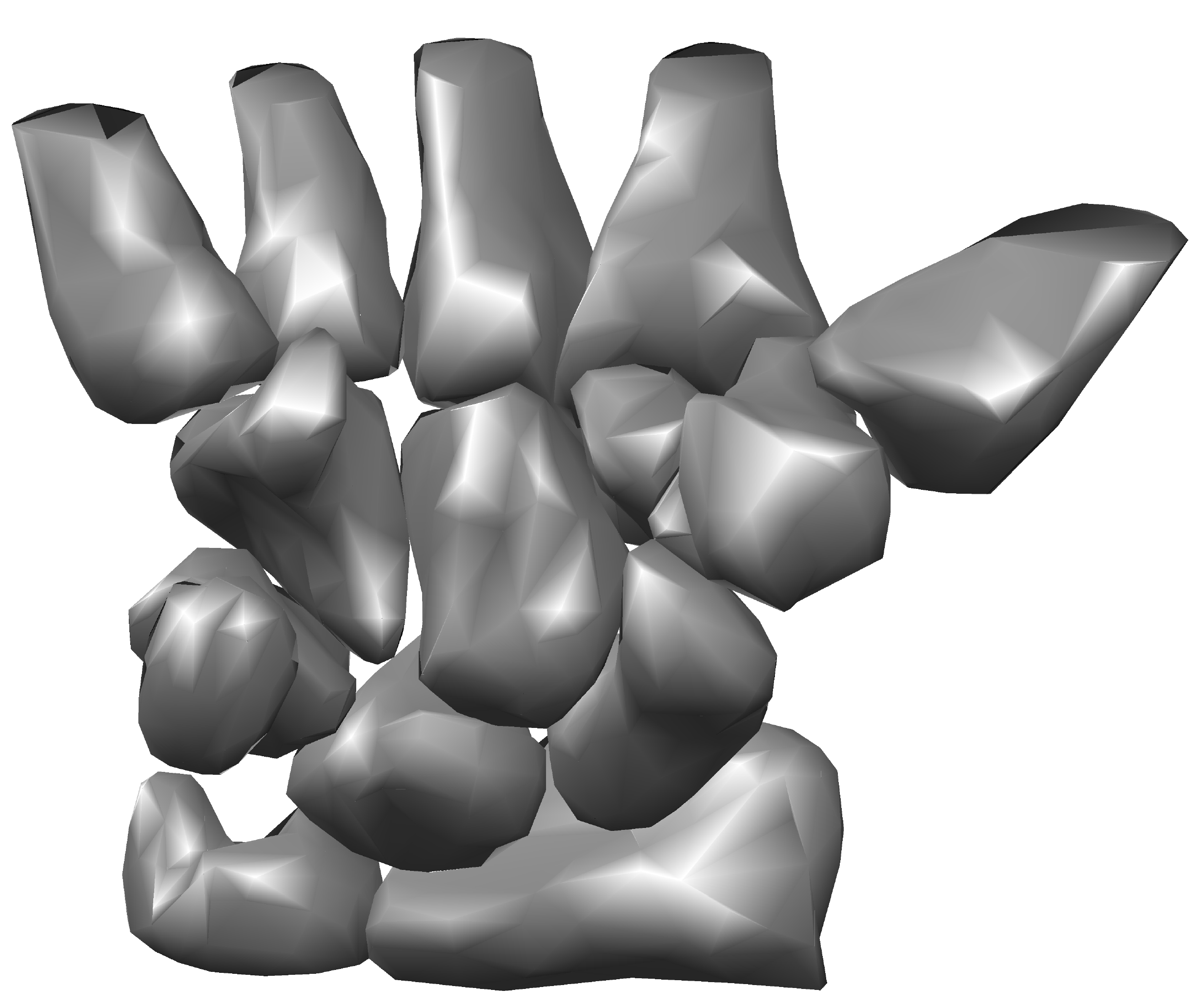}
\caption{3D representation of left carpal bone of a healthy male from palmar view.}
\label{Carpalbone}
\end{figure}

\medskip
\noindent{\textbf{Datasets}}\quad In order to evaluate the performance of the proposed GSGW framework on carpal bone surfaces a total of $20$ men and women with average age of $25$ years old from publicly-available benchmark \cite{Moore:07} have been chosen (see Figure \ref{Carpalbone}). More specifically, carpal bones of eight surfaces including capitate, hamate, lunate, pisiform, scaphoid, trapezium, trapezoid, triquetrum, and the first metacarpal for both wrists are used for performing analysis. Since the trapeziometacarpal joint of the thumb is a common site of osteoarthritis, the first metacarpal bone has been considered for our analysis. Each surface is rendered by triangular mesh with vertices and edges.

\medskip
\noindent{\textbf{Performance Evaluation Measures}}\quad
In a bid to compare the shapes of the carpal bones in women versus men, we generated the GSGW signature for each carpal bone (eight total) and the first metacarpal bone for each subject for both the right and left wrists. For fair comparison of the performance of our proposed GSGW approach with GPS embedding, we followed setting in \cite{Moore:07}. we also compared the GSGW for each bone of the right and left wrist separately for the two groups (ten women versus ten men) using one-way multivariate analysis of variance (MANOVA) based on sex. Furthermore, in order to to further confirm the excellency of our approach we exploited non-parametric permutation tests as the resampling procedure.

For permutation testing, gender labels of the samples are randomly shuffled for $1000$ times, and test statistics calculated the to generate the null distribution. Then, we created the same statistics without shuffling and compute the percentile of that score, which referred to as the $p$-value of the non-parametric permutation test. We report the $p$-value acquired from both MANOVA and permutation testing. For $p<0.05$, there would be a statistically significant difference between the two groups.

It should be noted that unlike GPS embedding in which the number of GPS coordinates varied for each carpal bone, we used a fixed dimension of $18$ which is acquired by computing principle component analysis of data matrix $\bm{X}$.

\medskip
\noindent{\textbf{Baseline Method}}\quad For the wrist benchmark \cite{Moore:07} used for experimentation, we will report the comparison results of our method against GPS embedding approach~\cite{Chaudhari:14}.

\medskip
\noindent{\textbf{Implementation Details}}\quad The experiments were conducted on a desktop computer with an Intel Core i5 processor running at 3.10 GHz and 8 GB RAM; and all the algorithms were implemented in MATLAB. The appropriate dimension (i.e. length or number of features) of a shape signature is problem-dependent and usually determined experimentally. For fair comparison, we used the same parameters that have been employed in the baseline methods, and in particular the dimensions of shape signatures. In our setup, a total of 31 eigenvalues and associated eigenfunctions of the LBO were computed. For the proposed approach, we set the resolution parameter to $R=30$ (i.e. the spectral graph wavelet signature matrix is of size $495\times m$, where $m$ is the number of mesh vertices).

\subsection{Carpal Bone Dataset}
Carpal bone dataset consists of $360$ mesh models from $20$ classes~\cite{Moore:07}. Each class contains $18$ objects with distinct postures. Moreover, each model in the dataset has approximately $m=1502$ vertices.

\medskip
\noindent{\textbf{Results}}\quad In our GSGW approach, each surface in the carpal bone dataset is represented by a $495\times 1502$ matrix of spectral graph wavelet signatures. The global spectral graph wavelet yields a $495\times 1$ vector $\bm{g}$ of spectral graph wavelet codes, resulting in a data matrix $\bm{X}$ of size $495\times 360$. Figure~\ref{Fig:GSGW1} shows the spectral graph wavelet codes of two carpal bones (capitate and lunate) from two different classes of carpal bone dataset. As can be seen, the global descriptors are quite different and hence they may be used efficiently to discriminate between surfaces in statistical analysis tasks.
\begin{figure}[!htb]
\setlength{\tabcolsep}{.5em}
\centering
\begin{tabular}{c}
\includegraphics{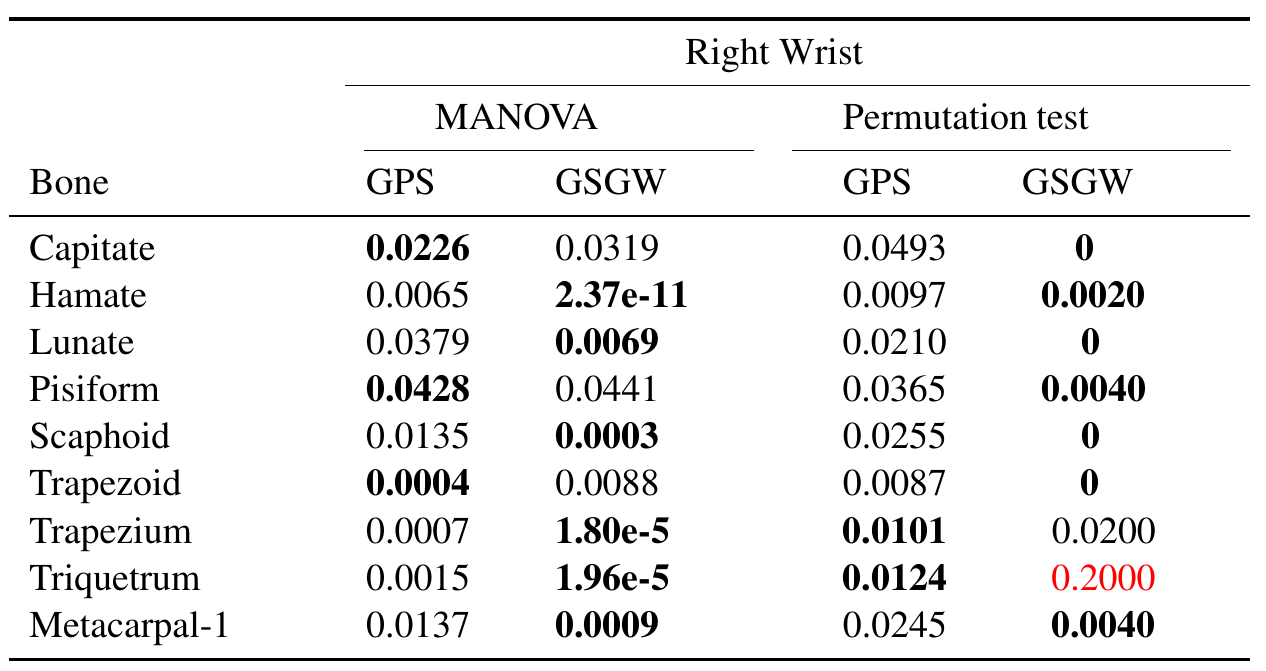}
\end{tabular}
\caption{Comparison of carpal bone surfaces of right wrist between males and females in terms of MANOVA and permutation test. Boldface numbers indicate the better performance while highlighted numbers show that the $p$-value is exceeded $0.05$}
\label{Fig:Right_Wrist}
\end{figure}

\bigskip

\begin{figure}[!htb]
\setlength{\tabcolsep}{.5em}
\centering
\begin{tabular}{c}
\includegraphics{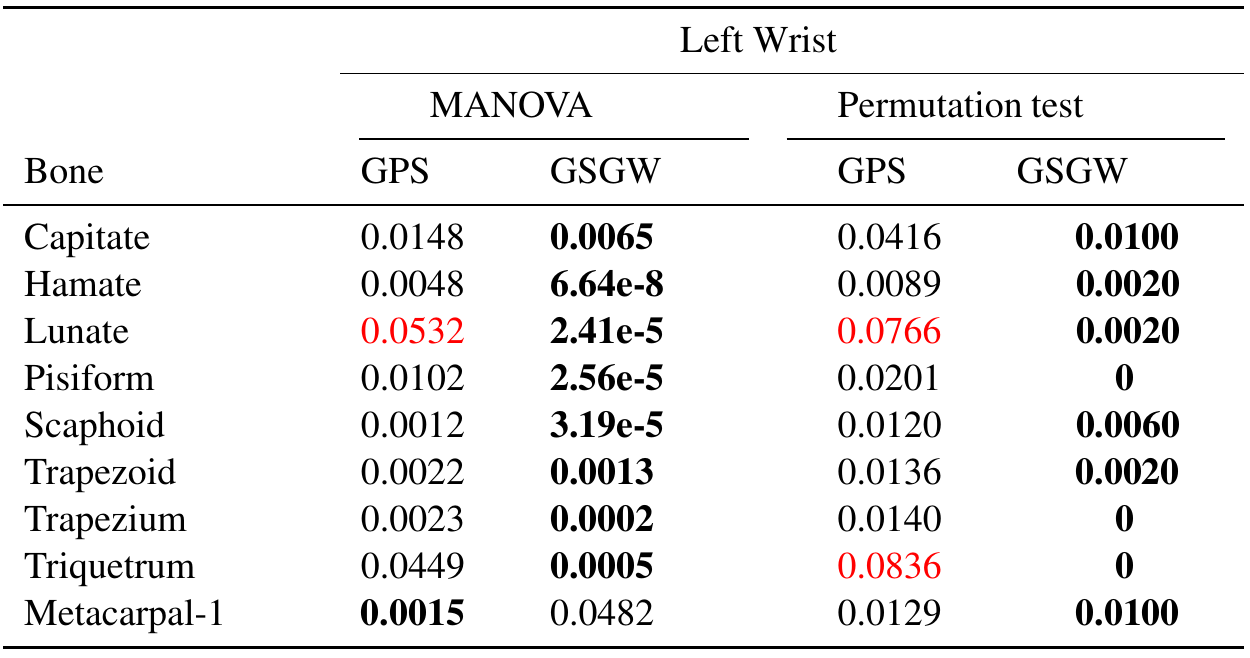}
\end{tabular}
\caption{Comparison of carpal bone surfaces of left wrist between males and females in terms of MANOVA and permutation test. Boldface numbers indicate the better performance while highlighted numbers show that the $p$-value is exceeded $0.05$}
\label{Fig:Left_Wrist}
\end{figure}
We compared our proposed GSGW method to GPS embedding method \cite{Chaudhari:14} in terms of MANOVA and non-parametric permutation test. The results are summarized in Table~\ref{Fig:Right_Wrist} for right wrist and Table~\ref{Fig:Left_Wrist} for left wrist, respectively. Highlighted numbers in Tables show that the $p$-value is exceeded $0.05$.

As can be seen, our method achieves better analytical performance than GPS embedding method for both right and left wrist. For right wrist, the GSGW approach yields the lower $p$-value compared with GPS embedding for six carpal bones out of nine ones in terms of MANOVA and for seven bones out of nine ones in terms of permutation test, respectively. In addition, the result of $p$-value of MANOVA for some bones, e.g. hamate, has improved up to $6.5 \times 10^{-3}$. For left wrist, our GSGW approach significantly improves the results by resulting the lower $p$-value in terms of MANOVA for all bones except Metacarpal-1. Also, our method achieved much lower  $p$-value in terms of permutation test for all carpal bones in contrast with GPS embedding. Moreover, the result of $p$-value of MANOVA for some bones, e.g. hamate, has improved up to $2.5 \times 10^{-3}$. To speed-up experiments, all shape signatures were computed offline, albeit their computation is quite inexpensive due in large part to the fact that only a relatively small number of eigenvalues of the LBO need to be calculated.

\begin{figure}[!htb]
\centering
\begin{tabular}{c}
\includegraphics[scale=.7]{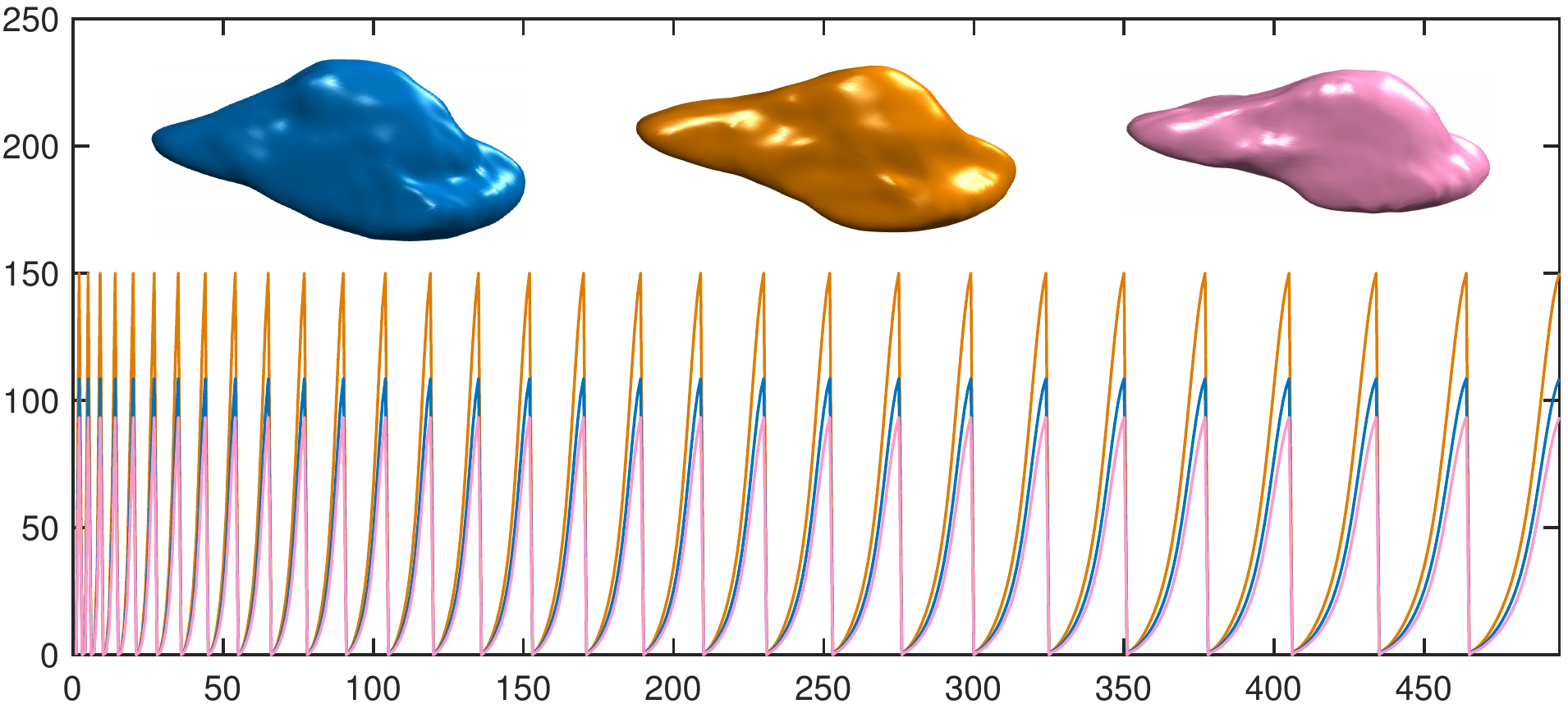}
\end{tabular}
\caption{Global spectral graph wavelet codes for three capitate bones belongs to women's left wrist.}
\label{Fig:GSGW2}
\end{figure}

\begin{figure}[!htb]
\centering
\begin{tabular}{c}
\includegraphics[scale=.7]{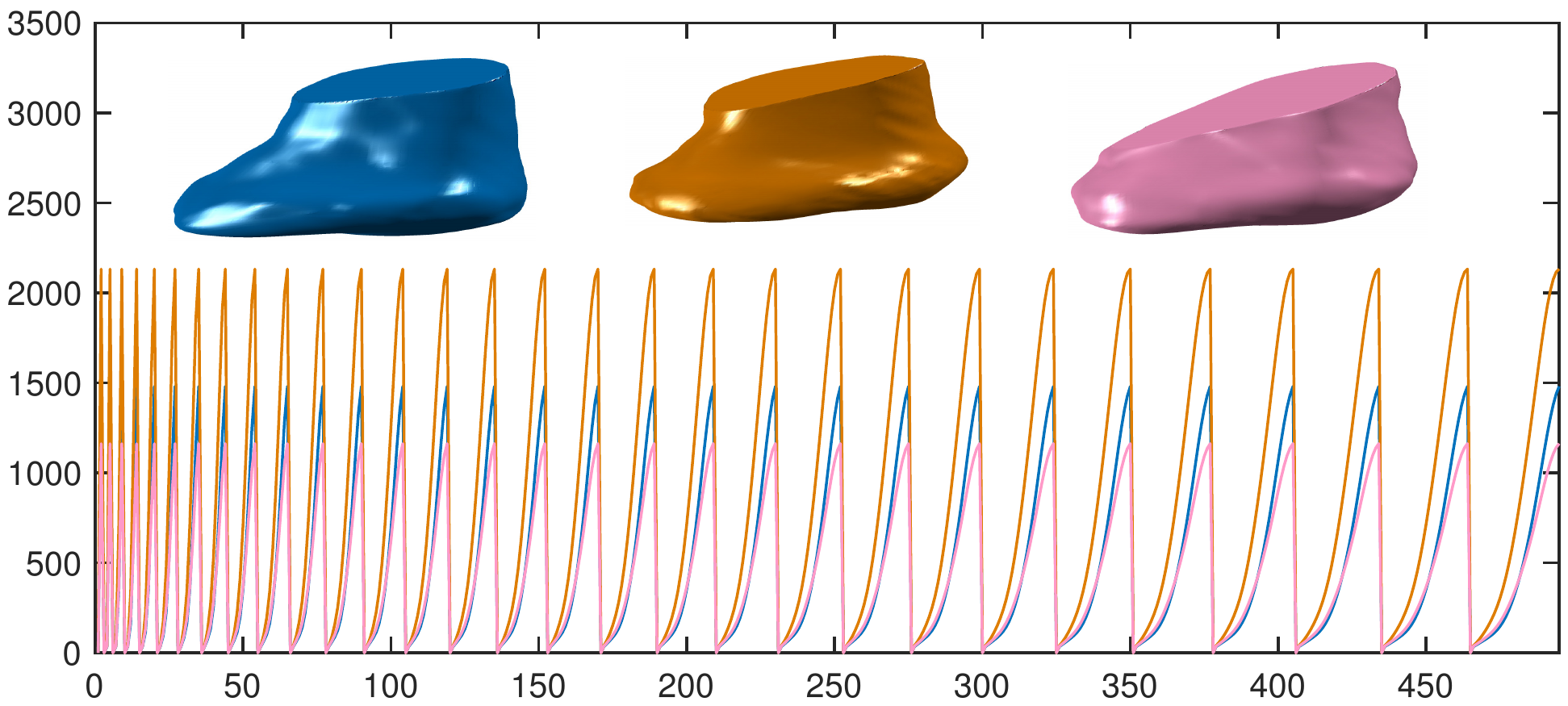}
\end{tabular}
\caption{Global spectral graph wavelet codes for three metacarpal bones belongs to women's left wrist.}
\label{Fig:GSGW3}
\end{figure}

In order to further assess the discriminative power of our approach, we computed the GSGW of carpal bone surfaces of the same class. As can be shown in Figures \ref{Fig:GSGW2} and \ref{Fig:GSGW3}, even for very similar carpal bones with slightly difference, the proposed GSGW is able to distinguish and recognize the shape.

\bigskip

\begin{figure}[!htb]
\setlength{\tabcolsep}{.5em}
\centering
\begin{tabular}{cc}
\includegraphics[scale=0.58]{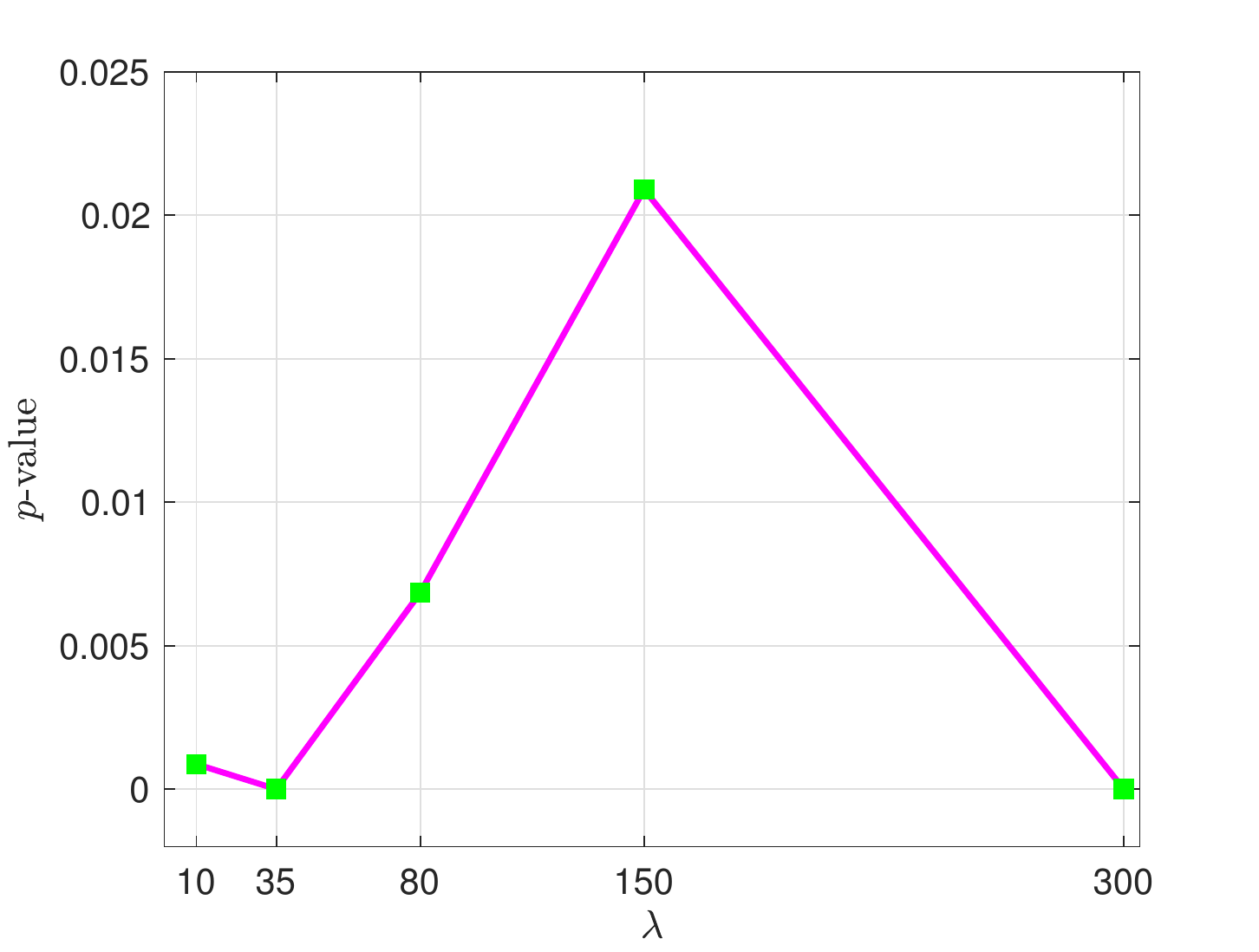} &
\includegraphics[scale=0.58]{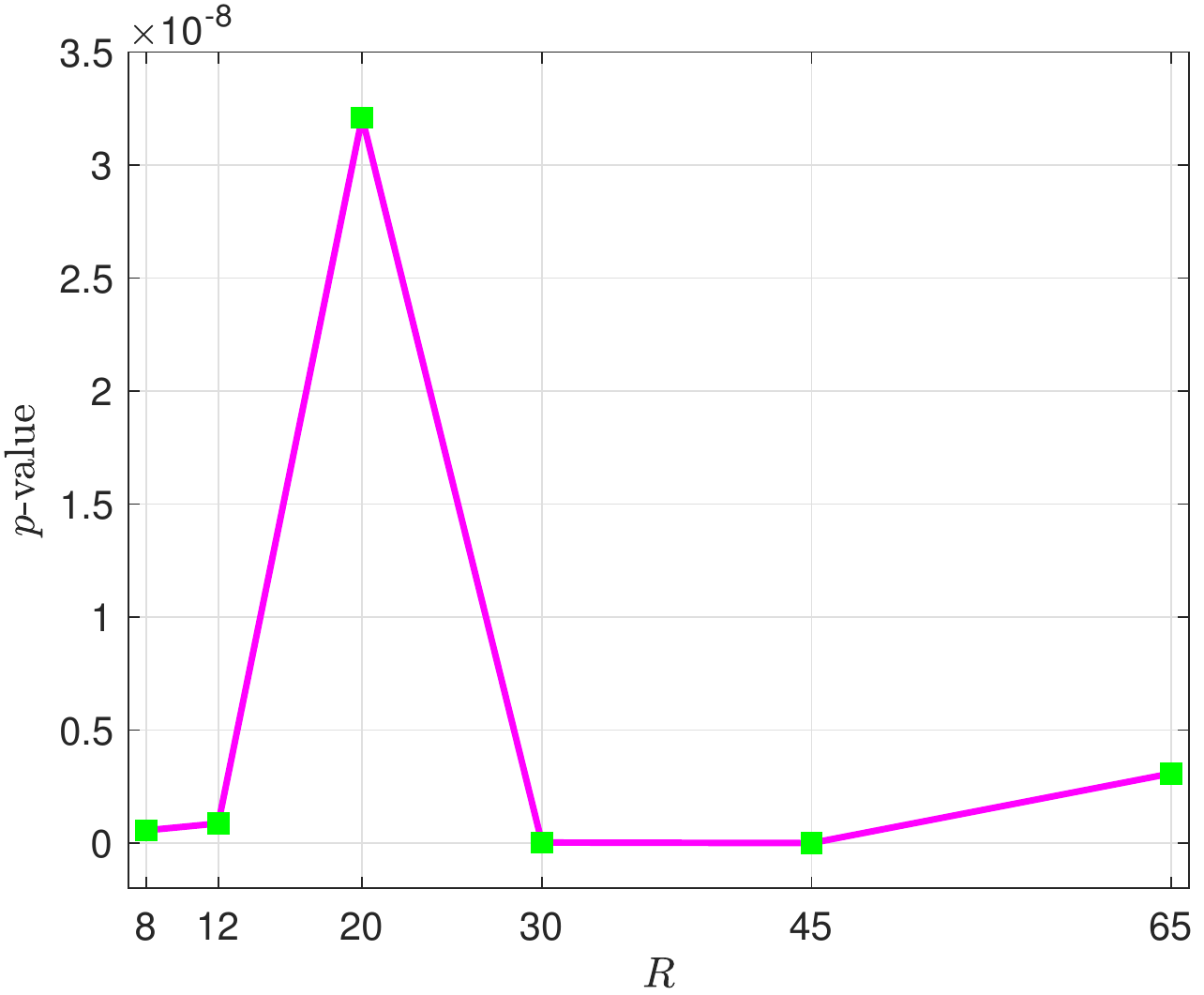}
\end{tabular}
\caption{Effects of the parameters on the MANOVA result for hamate bone.}
\label{Fig:parameters}
\end{figure}

\medskip
\noindent{\textbf{Parameter Sensitivity}}\label{PR}\quad
The proposed approach depends on two key parameters that affect its overall performance. The first parameter is the resolution $R$ of the spectral graph wavelet. The second parameter $\lambda$ is the number of eigenvalues of LBO, which plays an important role in the GSGW vector $\bm{g}$. As shown in Figure~\ref{Fig:parameters}, the best MANOVA result on carpal bones dataset is achieved using $R=30$ and $\lambda=30$. In addition, the performance of proposed method in terms of MANOVA is satisfactory for a wide range of parameter values, indicating the robustness of the proposed GSGW framework to the choice of these parameters.
\section{Conclusions} \label{conclusion}
In this paper, we presented a spectral-based framework for population study of carpal bones. Moreover, we performed statistical analysis on carpal bones of the human wrist by representing the cortical surface of the carpal bone using spectral graph wavelet descriptor to supply a means for comparing shapes of the carpal bones across populations. We also proposed a novel framework of directly extracting global descriptor so-called global spectral graph wavelet. Thus, we circumvented all the procedure of BoF paradigm which leads to a lower computation time as well as higher analysis accuracy. This approach not only captures the similarity between feature descriptors, but also substantially outperforms state-of-the-art methods both in accuracy and in scalability. For future work, we plan to apply the proposed approach to other surface analysis problems, and in particular segmentation and clustering.
\bibliographystyle{ieeetr}
\bibliography{biblio} 
\end{document}